\documentstyle[epsfig]{mn}

\newcommand{\acknowledgements}[1]{\vspace{7mm} \noindent {\normalsize \bf
Acknowledgements.\,} {\normalsize #1}}

\begin{document}

\title[Forming stars on a viscous timescale] 
{Forming stars on a viscous timescale: the key to exponential stellar profiles in disk galaxies?}

\author[A. Slyz et al.]  {Adrianne D. Slyz$^1$, Julien E. G. Devriendt$^2$,  
Joseph Silk$^2$, and Andreas Burkert$^1$\\
$^{1}$ Max Planck Institut f\"ur Astronomie, K\"onigstuhl 17, 69117 Heidelberg, Germany\\
$^{2}$ Nuclear and Astrophysics Laboratory, 
Keble Road, OX1 3RH Oxford, United Kingdom}

\maketitle

\begin{abstract}

We argue for implementing star formation on a viscous timescale in
hydrodynamical simulations of disk galaxy formation and evolution.
Modelling two-dimensional isolated disk galaxies with the 
Bhatnagar-Gross-Krook (BGK) hydrocode, we verify the analytic 
claim of various authors that if the characteristic timescale 
for star formation is equal to the viscous timescale in disks, 
the resulting stellar profile is exponential on several scale lengths 
whatever the initial gas and dark matter profile.
This casts new light on both numerical and semi-analytical disk formation
simulations which either (a) commence star formation in an already exponential
gaseous disk, (b) begin a disk simulation with conditions known to
lead to an exponential, {\em i.e.} the collapse of a spherically 
symmetric nearly uniform sphere of gas in solid body rotation under the
assumption of specific angular momentum conservation, or (c) in 
simulations performed in a hierarchical context, tune their feedback processes 
to delay disk formation until the dark matter halos are slowly evolving
and without much substructure so that the gas has the chance to
collapse under conditions known to give exponentials.
In such models, star formation  
follows a Schmidt-like law, which for lack of a suitable timescale, resorts 
to an efficiency parameter. 
With star formation prescribed  on a viscous timescale however,
we find gas and star fractions after $\sim$ 12 Gyr that are consistent 
with observations without having to invoke any ``fudge factor'' 
for star formation.
Our results strongly suggest that despite our gap in understanding the
exact link between star formation and viscosity, the viscous timescale
is indeed the natural timescale for star formation.

\end{abstract}


\section{Introduction}

The origin of the exponential radial light profiles in galaxy
disks is controversial.  The {\em{in situ}} scenario argues for stars forming
in an already
exponential gaseous disk (Gunn 1982; van der Kruit 1987;
Dalcanton, Spergel \& Summers 1997; Mo, Mao \& White 1998).  
The viscous scenario, on the other hand, contends that regardless of the initial gas 
density profile in the disk prior to star formation, the stars will 
acquire an exponential profile, provided star formation proceeds on a viscous
timescale (Silk \& Norman 1981; Lin \& Pringle 1987; Clarke 1989;
Yoshii \& Sommer-Larsen 1989; Saio \& Yoshii 1990; Olivier, Blumenthal
\& Primack 1991; Silk 2001; Ferguson \& Clarke 2001). 

The {\em{in situ}} scenario
defers the problem of the origin of exponential stellar light profiles to
the origin of the exponential gas distribution. 
The explanation for such a gas distribution stems from a remark made by Mestel (1963) 
on the similarity of the specific angular momentum distribution observed in the disks of 
spirals to that of a uniform sphere in solid body rotation. 
If specific angular momentum is conserved
during the collapse of such a gaseous sphere, the 
resulting disk naturally retains the sphere's original angular momentum distribution.
Considering the case where this uniform sphere of gas  
collapses within an isothermal dark halo,
Gunn (1982) found that the gas surface density distribution in the final 
disk would be approximately exponential over three scale lengths.

This simple model is the basis of recent explanations
of a host of disk properties.
By extending Gunn's scenario to
allow the dark halo to adiabatically respond to the accumulation of
baryonic mass in its center, Dalcanton {\em{et al.}} (1997) 
explained not only the structural properties of present-day disks
({\em{e.g.}} their exponential stellar profiles)
but also their dynamical properties ({\em{e.g.}}  their rotation curves, 
the Tully-Fisher relation). Using the Press-Schechter 
prescription to compute the abundance of virialized halos 
(Press \& Schechter 1974), combined with the prediction from numerical
simulations of both the halo dark matter density profiles
(Hernquist, 1990; Navarro, Frenk \& White 1996) 
and spin parameters 
(Barnes \& Efstathiou 1987; Warren {\em et al.} 1992),
these authors also predicted the surface brightness 
and scale length distributions of present-day disks.
Mo, Mao \& White (1998), using a similar modelling, extended this study 
to predict the evolution of disk properties within various cosmological models. 

In addition to the simple initial conditions set out by Mestel and Gunn,
models built upon the {\em in situ} star formation scenario require
a recipe for converting gas to stars. In the case of Dalcanton et al. (1997), 
Mo et al. (1998), and more generally in all semi-analytic models as well 
as numerical simulations, a Schmidt law,  with an efficiency 
parameter is adopted 
(see {\em e.g.} Kauffmann, White \& Guiderdoni 1993; Cole et al. 1994; 
Somerville \& Primack 1999; Katz 1992; Mihos \& Hernquist 1994;
Steinmetz \& M\"uller 1995; Gerritsen \& Icke 1997; Springel 1999).
Of course, to imprint the exponential profile 
on the stars, such a prescription must ignore any process that
could significantly redistribute the gas mass and angular momentum in the
disk, such as gaseous outflows or viscous flows induced by a turbulent 
ISM or spiral arms.  

Numerous hydrodynamical disk formation simulations have shed light on
 the {\em in situ} theory for the origin of exponential disk profiles.  
The earliest
were N-body plus SPH simulations, first without
star formation (Katz \& Gunn 1991) and then with stars
(Katz 1992).  These simulations followed the collapse of an
isolated, uniform density perturbation in solid body rotation embedded
in an expanding universe.  External tidal fields were neglected. 
Since some small-scale power was added, dark matter and gas clumped
prior to the gas dissipating its energy and falling to form a disk.
Obviously, the merging of these subclumps involved angular momentum transfer,
but apparently the deviation from Mestel and Gunn's idealized
initial conditions was minor and the final gaseous and
stellar profiles were well-fit by exponentials.
All other simulations starting from similar initial conditions have
led to the same conclusion (Vedel, Hellsten \& Sommer-Larsen 1994; Steinmetz \& M\"uller 1995;
Contardo, Steinmetz \& Fritz-von Alvensleben 1998).

In the current paradigm of hierarchical galaxy formation however, one needs
to follow the formation of disk galaxies in the context of a more realistic tidal
field, for it seems implausible in such a paradigm that all gas destined to settle into a
disk  originated in a rigidly rotating spherical protocloud
which conserved its specific angular momentum not only during
the process of protogalactic collapse and settling into 
centrifugal equilibrium but also throughout the star formation process.  
Unsurprisingly, hydrodynamical simulations of hierarchical galaxy
formation illustrate how nonspherical subclumps of
gas and dark matter torque eachother while merging. In fact,
in such simulations of only gas and dark matter, there is so much 
transfer of angular momentum during merging that the
disks which form have angular momenta two orders of magnitude lower than
what is observed (Navarro \& Steinmetz 1997).

To remedy this, attempts have been
made to delay the cooling and subsequent
collapse of gas in dark halos, so that disks could form  late in cosmological simulations when
dark matter halos are evolving more slowly and have much less
substructure (Weil, Eke \& Efstathiou 1998). In these situations,
the idealized initial conditions envisioned by Gunn (1982)
may be an accurate starting point for disk formation.  
First attempts at physically motivating such a delay by
including feedback mechanisms instead of just switching on cooling
in a simulation at a late epoch have been undertaken by Sommer-Larsen, 
Gelato \& Vedel (1999).  In their most successful cases, the gas 
surface densities of their disks are poorly represented by exponentials (their Figure 4). 
However, resolution in their simulation (between 3303 and 5017
gas particles in a disk) is such that determining a disk scale length is already a difficult task. 
Hence it may be resolution that prevents these profiles from looking more
exponential. At any rate, on the evidence of current numerical simulations,
we find it hard to be convinced that all or even most gaseous disks forming
in a hierarchical context will acquire an exponential profile.

Rather than trying to fine tune the simulations to obtain idealized initial
conditions, it seems more appealing to look for a mechanism that could drive the stellar 
profile towards an exponential, regardless of the gas profile it starts from. 
Bearing in mind that star formation and the accompanying complexity of stellar
feedback are poorly understood and would be hard to simulate (for obvious resolution issues) 
even if they were understood, the best that one can do is to model processes which govern star
formation on a large scale. 
With this intention, as an alternative to Katz's (1992) star formation recipe based on
the physics of Jeans instability, we propose to capture
the physics that drives star formation by modelling
the interstellar medium as a turbulent fluid of colliding, fragmenting 
and star-forming clouds with energy to drive the turbulence supplied from supernovae, stellar
winds, and ionizing radiation from newly born stars.
Such turbulence would drive viscous flows.

Therefore, as a start, in this paper, we use numerical hydrodynamical simulations
to revisit the hypothesis that star formation occuring on a viscous timescale 
in galaxy disks yields exponential stellar profiles. Lin \& Pringle (1987) 
explored this question by numerically integrating 1D differential equations, and
Yoshii \& Sommer-Larsen (1989) confirmed their result by deriving analytical solutions for
the star forming viscous disk. Saio \& Yoshii (1990) extended these
works by including self-gravity.
The numerical hydrodynamical approach finds its justification in the fact that, in the future,
we want to extend these investigations beyond 1D calculations, and with physics that is 
difficult to include in an analytic approach ({\em{e.g.}} stellar mass loss, supernovae feedback).

The outline of the paper is as follows: 
after a discussion of turbulence as a plausible source of viscosity in galaxy disks
(section~\ref{viscous_turb}) and a description of our numerical 
approach (section~\ref{num_approach}), we commence with benchmarking 
our code to the viscous evolution of a pure gaseous disk without star 
formation (section~\ref{visc_pure_gas}).  Then we proceed to a 
detailed examination of 
experiments of star formation in viscous galaxy disks with different 
initial gas profiles and dark halo rotation curves 
(section~\ref{SF_in_visc_disks}). Finally, we discuss our results 
(section~\ref{conclusions}).

\section{Physical motivation for viscosity}
\label{viscous_turb}
A general consensus among the astrophysical community is that the
true kinematic viscosity of the interstellar medium, $\nu$, is 
much too low to alter the galactic disk structure on a timescale comparable to the 
age of the universe.
It has been argued by various authors (Lynden-Bell and Pringle 1974;
Silk and Norman 1981; Lin and Pringle 1987; Yoshii and Sommer-Larsen 1989;
Clarke 1989; Olivier et al. 1991; Duschl et al. 2000; Silk 2001) however, that
viscous turbulence could drive a non-negligible radial flow of gas in 
galactic disks.

Turbulence is a consequence of fluid motion, therefore to be maintained it
requires an external energy source $E_{ext}$.
This energy which arises on large scales, with characteristic 
velocity $U$ and length $L$,
is progressively transferred to smaller and smaller scales
by non-linear interactions, until the dissipation scale $L_{dis}$ is 
reached where molecular processes convert it into heat.
It follows from dimensional arguments 
that $E_{ext} \sim U^3/L$. Similarly, one can obtain
$L_{dis} \sim (\nu^3 E_{ext}^{-1})^{1/4}$.
As a result, 
\begin{eqnarray}
\frac{L}{L_{dis}} = {\mathrm{Re}}^{3/4} \,\, , {\mathrm{Re}} = \frac{U L}{\nu}
\label{eqdis}
\end{eqnarray}
where ${\mathrm{Re}}$ is the Reynolds number. 

Although star formation and/or magnetic fields certainly participate in feeding turbulence, 
we may suppose that turbulence is driven by differential rotation, (as argued by {\em e.g.} 
Richard and Zahn (1999) for accretion disks), for
the following qualitative argument does not change.
Typical values for a galactic disk are $L \sim 10^{19}$m (radius
of an orbit), $U \sim 10^5 \rm{m~s}^{-1}$ (corresponding azimuthal velocity) and 
$\nu \sim 10^{-11} \rm{m}^2~\rm{s}^{-1}$ (we have assumed an average density of 
$10^6 \rm{at~m}^{-3}$, a temperature of $10^4$K and described the gas 
as elastic spheres of diameter $10^{-10}$m). Combined together, these numbers lead to 
${\mathrm{Re}} \sim 10^{35}$. This extremely high Reynolds number means
that in order to fully resolve our problem numerically  
in 3D, we should have about $10^{79}$ grid points (see eq.~\ref{eqdis}), 
whereas state-of-the-art super computers can handle $10^{10}$ at most. 
Therefore, the only way around this problem is to somehow model what happens on subgrid scales 
and numerically resolve the properties of the flow on large scales. 

As first noted by Boussinesq (1870), it seems that to first order
and on macroscopic scales the net effect of turbulence is simply
to turn the real fluid into a more viscous one.  
This can be schematically understood as follows: the Reynolds number of 
a turbulent fluid is so high 
that any perturbation in the smooth flow produces eddies which  
dissipate energy until the flow reaches a critical, lower Reynolds number 
$\rm{Re}_c$, corresponding to the onset of turbulence. If more eddies 
are produced, the effective viscosity  becomes too high, cutting 
off the production of eddies. On the other hand, 
if not enough eddies are produced, then the flow  still exhibits 
a tendency to produce turbulence.   
Laboratory flow values for critical Reynolds numbers 
are on the order of $\rm{Re}_c \sim 10 - 10^3$
(see for example Richard and Zahn 1999), so that one can
solve the Navier--Stokes problem with this critical Reynolds number 
and hope that it is enough to obtain a good description of the properties of the 
flow on macroscopic scales.

Our goal here is not to derive a mathematically rigorous formulation 
of hydrodynamic turbulence in galactic disks but to suggest that there 
exist strong physical arguments in favor of describing the behaviour 
of the gas flow on large scales by simply solving a Navier--Stokes problem 
with a smaller Reynolds number, 
namely the critical Reynold's number, $\rm{Re}_c \sim 10 - 10^3$.

\section{Numerical approach}
\label{num_approach}
We perform numerical hydrodynamical simulations in 2D with the BGK 
(Bhatnagar-Gross-Krook) code (Prendergast and Xu 1993, 
Slyz and Prendergast 1999). This
is a scheme based on gas-kinetic considerations, meaning the
code solves for the time evolution of a gas distribution
function, $f({\vec x},{\vec u},t)$, in phase space. Fluxes are
computed from velocity moments of $f$. 
$f$ evolves in time according to an approximation to the 
collisional Boltzmann equation, namely
the BGK equation: $\frac{Df}{Dt} = \frac{g - f}{\tau} $
%\begin{eqnarray}
%\frac{\partial f}{\partial t}  +  {\vec u}\cdot \frac{\partial f}{\partial {\vec x}}
%-  \frac{\partial{\Phi}}{\partial {\vec x}} \cdot
%\frac{\partial f}{\partial {\vec u}} = \frac{g - f}{\tau}.
%\label{eq:cylBGK}
%\end{eqnarray}
where $\frac{D}{Dt}$ 
is a time rate of
change along the  trajectory of a single  particle 
moving freely in phase space under the action of a smoothly varying
gravitational field and where $\frac{g - f}{\tau}$ models
the effects of collisions. Here $g$ is a local equilibrium distribution, 
{\em i.e.} the Maxwell-Boltzmann distribution function, having the same mass, 
momentum and energy densities as $f$, and $\tau$ is a relaxation time, 
which can  be as small as the mean time between collisions of 
a particle in the gas (Bhatnagar, Gross \& Krook 1954).  The 
underlying hypothesis of the BGK model is that  there is a
relaxation process in which
collisions tend to reshuffle the particles defining the true 
distribution function, 
$f({{\vec x}}, {{\vec u}},t)$ to
$g({{\vec x}}, {{\vec u}},t)$,  while conserving the total mass, momentum
and energy.  This conservation requires that the moments of the BGK collision 
term must vanish.  
$f$'s relaxation to the equilibrium state $g$ is dissipative and hence
is accompanied by an increase in entropy.  It is nontrivial that the BGK
scheme captures this physics. Many schemes have to invoke an ``entropy fix''
usually in the form of artificial viscosity to mimic an entropy increase
and eliminate unphysical numerical phenomena such as rarefaction shocks.

Together with the criteria of vanishing moments of the collision term,
a numerical solution of the formal integral  solution (eq.~\ref{eq:formalBGKsoltn} below) of the BGK
equation comprises the BGK scheme.  
Given initial conditions $f =
f_{0}$ at $t = t_{0} = 0$, the solution to the BGK equation
contains two terms: an integral over the past
history of $g$ and a term representing relaxation from an initial
state, $f_{0}$ on a timescale $\tau$.  
\begin{eqnarray}
f({\vec x}_{f},{\vec u},t)=\frac{1}{\tau} \int^t_{0}
g({{\vec x}}^{'},{{\vec u}}^{'},
t^{'}) e^{-(t-t^{'})/\tau} \, dt^{'} + \nonumber \\
e^{-t/\tau} \, f_{0}({\vec x}_{f} - {\vec u} t,{\vec u},t_{0})
\label{eq:formalBGKsoltn}
\end{eqnarray}
Here ${\vec x}^{'}$ and ${\vec u}^{'}$ in the arguments of $g$ are 
points along a gas particle's trajectory in phase space
with the final conditions ${\vec x}^{'} = {\vec x}_{f}$ and
${\vec u}^{'} = {\vec u}$ at $t^{'}=t$.
When fluxes are eventually 
computed at ${\vec x}_{f}$, $f({\vec x}_{f},{\vec u},t)$  
is integrated over the full and continuous
range of velocities, ${\vec u}$, from $-\infty$ to $+\infty$.  
Therefore fluxes computed from the
true distribution function $f({\vec x}_{f}, {\vec u}, t)$ arise
by a weighted integration of the equilibrium distribution function
$g({{\vec x}^{'}}, {{\vec u}^{'}}, t^{'})$  not just over one trajectory but 
over all possible trajectories (in 6-dimensional phase space)
which arrive at (${\vec x}_{f},t$).  A detailed description
of the numerical procedure in the BGK scheme for the
2D Cartesian case is given in Xu (1998).

By evolving $f$ through an equation which accounts for particle collisions, 
the fundamental mechanism  for generating dissipation in gas flow, 
the BGK scheme gives fluxes which carry both advective and dissipative terms. 
This means that for viscous problems no modification to the 
BGK flux expressions from their form for non-viscous problems
is necessary.   Furthermore there is no need for 
viscous source terms. More than
this, unlike the earlier gas-kinetic scheme used in
astrophysics, namely the beam scheme (Sanders and Prendergast 1974),
which also carried dissipation in its flux expressions, the BGK scheme
is dissipative in a realistic way. Where the beam scheme arbitrarily
chooses the form of $f$ at the beginning of each updating time step 
and evolves it through the collisionless
Boltzmann equation, the BGK scheme solves for the time evolution of $f$
throughout an updating time step using a model
of the collisional Boltzmann equation. 
If the gas happened to be in disequilibrium
at the beginning of the updating time step, the beam scheme
cannot readjust the gas properly to equilibrium because it evolves $f$
through the collisionless Boltzmann equation which is simply
missing the adjusting mechanism provided by collisions. 

Also by assuming instantaneous relaxation back to the
chosen form for $f$ at the beginning of an updating time step, the
beam scheme endows the gas with a mean collision time equivalent
to the updating time step. Collisions in the BGK scheme on the other hand
are active throughout the updating time step and for hydrodynamical 
applications the BGK scheme demands that the collision time is much smaller 
than the updating time step. Since dissipation parameters are proportional
to the collision time, $\tau$, {\em e.g.} the dynamical viscosity $\eta = \tau p$, 
and the heat conduction coefficient 
$\kappa = \tau \frac{K+5}{2} \frac{p k}{m}$ (where $p$, $k$, $m$ are
the gas pressure, Boltzmann constant, and particle mass respectively
and $K = \frac{-3 \gamma +5}{\gamma -1}$ where $\gamma$ is the ratio
of specific heats, $\frac{c_p}{c_v}$), we easily see that an overestimation
of the collision time will lead to a very diffusive scheme.

The BGK code has been extensively
tested on standard 1D and 2D test cases of discontinuous nonequilibrium flow
with features ranging from cavitation
(Roe test) to the collision of strong shocks (Woodward-Collela test)
(see Xu 1998 for a review).
At shocks and contact discontinuities the code behaves as well 
as the best high resolution codes which do not employ regridding in
the neighborhood of a shock front and it gives better results at 
rarefaction waves.  Because the BGK scheme naturally satisfies the
entropy condition, it never requires an entropy fix when it encounters
strong rarefaction waves (cf. the Sj{\"{o}}green test). Tests of the
BGK schemes ability to solve the Navier-Stokes equations in smooth
flow regions include the Kolmogorov and laminar boundary layer
problem (Xu and Prendergast 1994). Tests of the BGK scheme in
an external gravitational potential show the long-term stability
of the scheme and its convergence to the equilibrium solution
(Slyz \& Prendergast 1999).

\subsection{Quantifying Viscosity in BGK}
\label{text_quant_visc}
As already mentioned in the previous section, on a microscopic
level dissipative parameters are directly related to 
the collision time of the gas, $\tau$.  Like other finite volume
schemes, the BGK scheme follows the mass, momentum and
energy densities within cells on an Eulerian grid.
For the flux computation, the BGK scheme solves for $f$ in the
neighborhood of the boundary of each computational cell and
for a short time (given by the usual CFL (Courant-Friedrichs-Lewy) condition).
Hence at each time step and at each cell boundary, it is possible
to solve the BGK equation with the collision time $\tau$ which is 
appropriate to the local fluid properties. For example, since gas 
kinetic theory gives a dependence of collision
time on gas density $\rho$ and temperature $T$, an expression for $\tau$ 
may be written as $\tau = \frac{{\cal C}_{1}}{\rho\sqrt{T}} $ 
where ${\cal C}_{1}$ is a proportionality constant which is chosen 
according to the desired Reynolds number of the problem. 
(${\cal C}_{1} = (UL/Re)(\rho^2/\sqrt\lambda p)$ where
U and L are characteristic velocity and length scales, $\lambda = \frac{m}{2kT}$ where $T$ is the gas temperature, $p$ is the gas pressure, and $Re$ is
the Reynolds number.) If we wish
to do a problem with constant kinematic viscosity for the case
where the Reynolds number is equal to the critical Reynolds number, $Re_c$, and
an isothermal gas, then we set $\tau$ to have a constant value,
$\tau = \frac{2 \lambda U L}{Re_c}$.
Once $\tau$
is known, we can write an expression for the real physical 
kinematic viscosity coefficient $\nu = \tau \frac{p}{\rho}$.

For problems in which the gas flow has discontinuities, the collision
time $\tau$ serves an additional purpose. If the grid is not fine enough
to resolve a discontinuity, then artificial dissipation must be added
to broaden the discontinuity so that it is
at least one grid cell thick.  Because viscosity and
heat conductivity are proportional to $\tau$, the BGK scheme broadens 
shocks by enlarging $\tau$ at the location of the discontinuities.  
The expression for the collision time in the BGK scheme therefore
contains another term in addition to the real physical collision time. 
The second term is chosen in such a way that shocks in the flow span
at least one grid cell.  In 
one-dimensional Cartesian coordinates this second term is:
\begin{equation}
{\cal C}_{2}
\frac{\arrowvert p_{l} - p_{r}  \arrowvert}
{(p_{l}  + p_{r})}.
\label{eq:secondcollterm}
\end{equation}
The subscript $l\hspace{.3em}(r)$ denotes quantities interpolated from the
left (right) side of the cell interface at which we compute
the collision time $\tau$, and $p$ is the gas pressure.

This second term tunes the amount of artificial
dissipation in the scheme.  The notable difference between how the
BGK scheme inputs artificial dissipation and how other schemes input
artificial dissipation is that the BGK scheme puts it in exactly as
if it were real dissipation corresponding to the numerically necessary
value for $\tau$. We again emphasize that neither the artificial dissipation
nor the physical dissipation enter the scheme as a source term.
When the second term (eq.~\ref{eq:secondcollterm}) in the collision time
dominates,  the order of the scheme is reduced and the true
distribution function $f$ is determined more by the initial
distribution function $f_{0}$ than by the integral over $g$ (eq.~\ref{eq:formalBGKsoltn}).
For smooth flow however, such as is encountered in the viscous
evolution of a galactic disk in an axisymmetric gravitational potential,
$C_2$ may be set to zero and the BGK scheme evolves
the gas with its true physical dissipation parameters, as long as the
grid resolution corresponding to the desired Reynolds number of
the problem is achieved.

Even when $C_2$ is set to zero, in addition to the (explicit) real
physical viscosity in the scheme  (controlled by the collision time) there is 
an inevitable numerical  viscosity, dependent on the grid spacing: 
the coarser the grid, the larger this numerical viscosity. 
The origin of this viscosity is that the scheme forgets 
everything except cell averages at the end of each time step. 
A consequence of this is that when one does a resolution study
of the viscous evolution of a gaseous disk, the flow properties
on a 207 X 207 grid quantitatively match those on the 85 X 85 grid 
only when one runs the problem on the finer grid with larger
$\tau$.

%We remark that at a given time $t$ the BGK
%collision term $\frac{( g - f)}{\tau}$ assigns the same collision 
%time $\tau$ to all particles in a given volume element of phase space, 
%regardless of their velocity.
%This simplification restricts the BGK scheme so that it may
%recover  Navier-Stokes solutions  only for flows with fixed 
%Prandtl number of 1.  This simplification also accounts for why $\tau$ 
%appears outside of the integral over time $t^\prime$ in the  
%formal solution for $f$ (equation~\ref{eq:formalBGKsoltn}).  

\begin{figure}
\centerline{\psfig{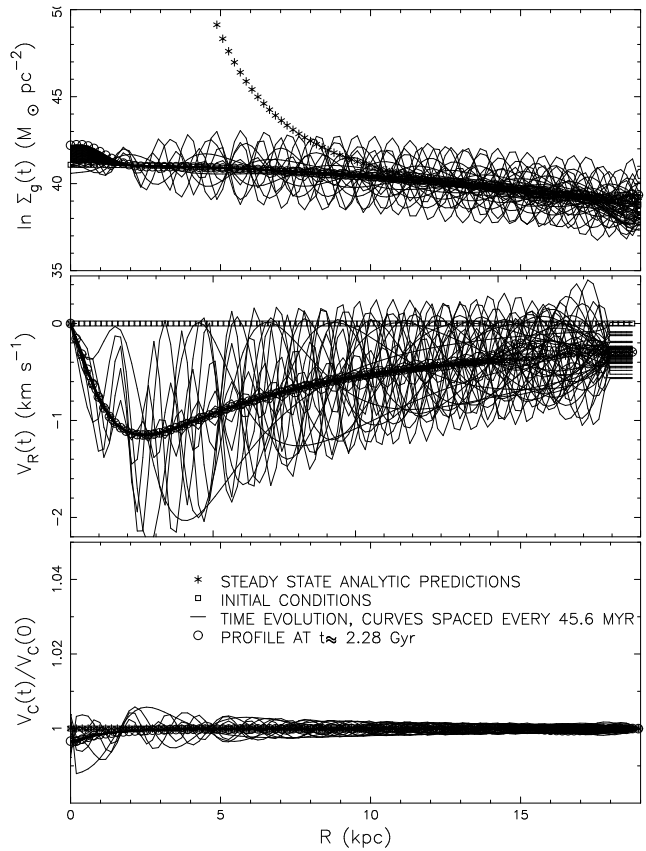}}
  \caption[]{Early viscous evolution (first 2.28 Gyr) of a disk of pure gas
initially in inviscid centrifugal equilibrium ($v_{R}=0$)
in a fixed gravitational potential. Coding for the different curves 
is given in the bottom panel of the figure.}
  \label{gasprof_early_zero}
\end{figure}

\section{Viscous Evolution of a Disk of Pure Gas}
\label{visc_pure_gas}

Before testing the hypothesis that $t_\star \sim t_{visc}$ gives
exponential stellar profiles, we check
how closely the BGK hydrocode reproduces our analytic
predictions for the steady state viscous evolution of a gas without star
formation.  The interest of performing such a computation is obvious: 
it allows us to measure the true viscosity of the code as well as its 
ability to capture the Navier-Stokes solution for a compressible viscous 
fluid evolving in a fixed gravitational field for the case
where the shear flow is highly supersonic (Mach $\approx$ 20) throughout
most of the disk.  Earlier tests of BGK's ability to give
Navier-Stokes solutions were only for subsonic shear flows
(Mach $\approx$ .5 for the laminar boundary layer problem;
Mach $\approx$ .8 for the decaying sinuisoidal wave problem)
(Xu and Prendergast 1994).

We initialize the gas in a fixed dark halo with gravitational potential:
\begin{eqnarray}
\Phi(r) = \frac{1}{2} {{\rm v_0}}^2 {\rm ln}({{\rm r_0}}^2 + {\rm r}^2)
\label{eq:logpotential}
\end{eqnarray}
where $v_0 = 220$ km/s  and $r_0=2.56$ kpc.
We consider two cases: the case where the gas is initially 
in inviscid centrifugal
equilibrium ($v_{R}=0$), and the case where the gas is in centrifugal
equilibrium but the initial radial velocity
profile is derived from the Navier-Stokes equations (instead of from
Euler's equations) given our choice
of initial gas profile (see Appendix A). We neglect self-gravity and
assume constant kinematic viscosity, $\nu$, and uniform
temperature at all times ($T \approx 10^4$ K).  We comment more on
the astrophysical motivation behind our choice of initialization in
section~\ref{initial_bc}.

To obtain a 
constant $\nu$ in the computation, the 
collision time $\tau$ in the BGK scheme is set to
$\frac{2 \lambda U L}{Re_c}$ (see discussion in section~\ref{text_quant_visc}) where $U= 220$ km/s, $L=20$ kpc,
$Re_c \approx 750$.
The computation is performed on a 207 X 207 Cartesian grid and
we follow the flow for about 12 Gyr. We would like to point out that in order 
to preserve the cylindrical symmetry of the flow on such a grid we had to use the high 
order limiter described in Huynh (1995).
We specify boundary conditions in detail in section~\ref{initial_bc} below.

\begin{figure}
\centerline{\psfig{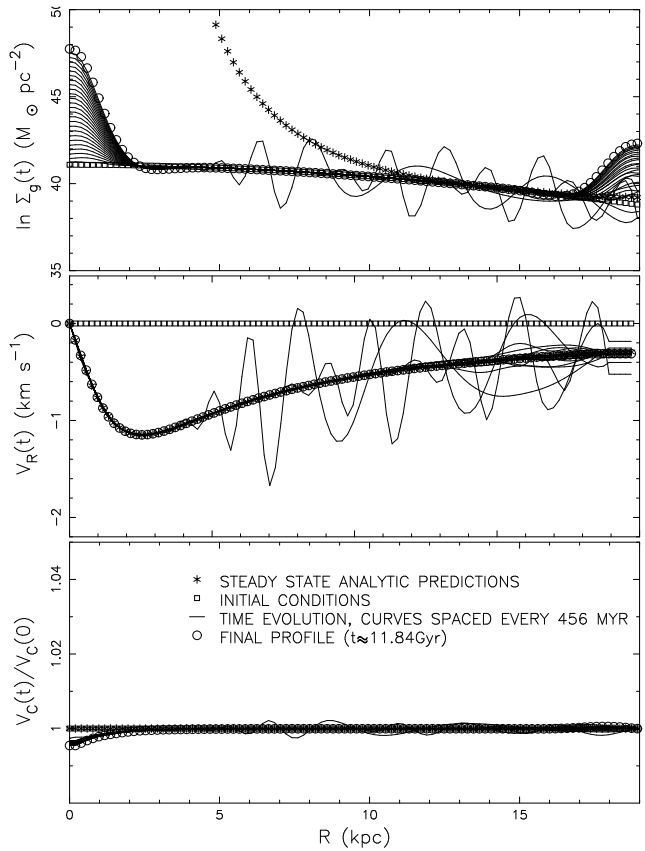}}
  \caption[]{Same run as in fig.~\ref{gasprof_early_zero} but with the results for
the evolution shown up to 11.84 Gyr. The steady-state analytic predictions (asterices)
are compared to the BGK numerical results (thin solid curves and empty circles).}
  \label{gasprof_zero}
\end{figure}

\begin{figure}
\centerline{\psfig{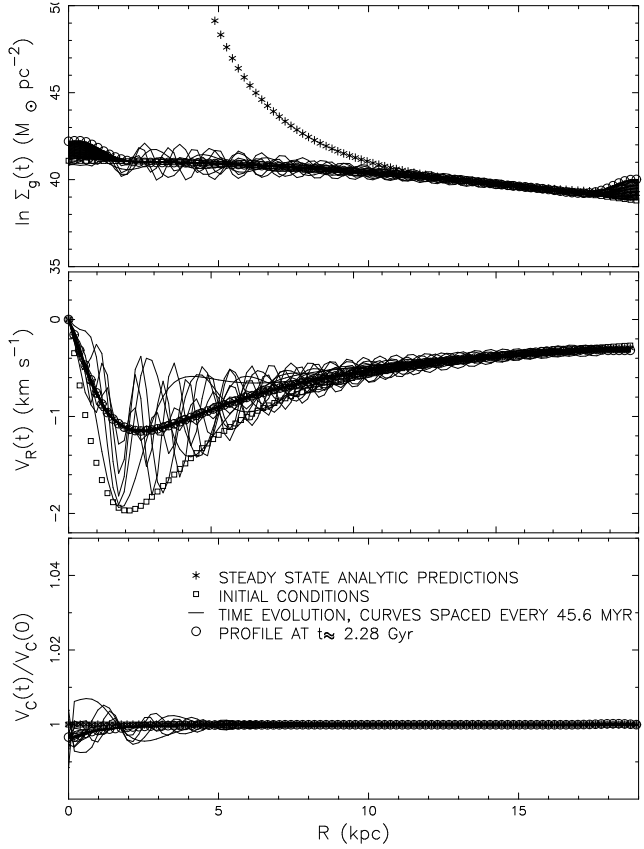}}
  \caption[]{
Early viscous evolution (first 2.28 Gyr) of a disk of pure gas
in a fixed gravitational potential. This time the initial radial velocity is
set to the velocity profile given by the solution to the Navier-Stokes equation
corresponding to the initial density profile (see Appendix A for details). Coding for 
the different curves is given in the bottom panel of the figure.}
  \label{gasprof_early}
\end{figure}

\begin{figure}
\centerline{\psfig{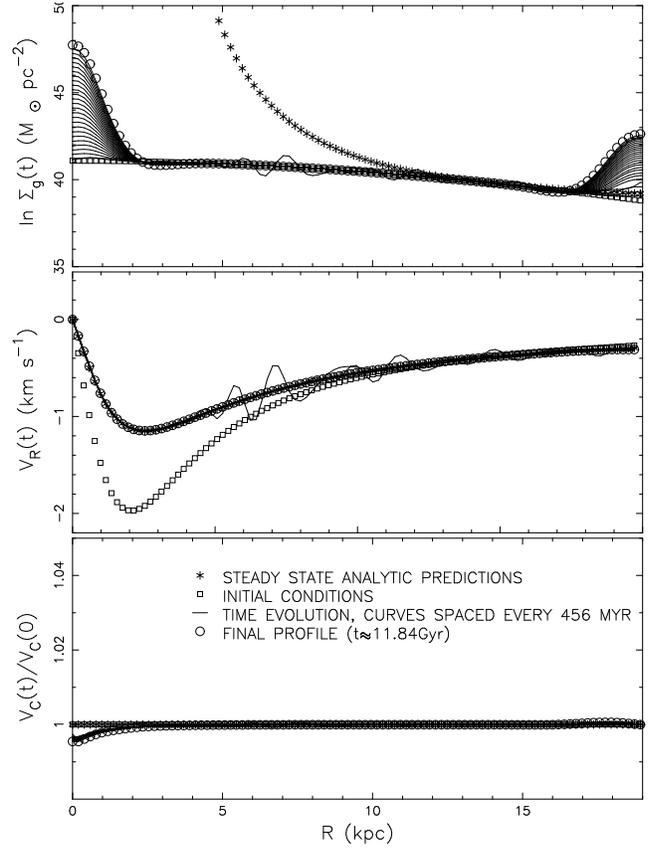}}
  \caption[]{Same run as in fig.~\ref{gasprof_early} but with the results for
the evolution shown up to 11.84 Gyr. The steady-state analytic predictions (asterices)
are compared to the BGK numerical results (thin solid curves and empty circles).
}
  \label{gasprof}
\end{figure}

In figures~\ref{gasprof_early_zero} and~\ref{gasprof_zero} we plot the code's results for the time evolution 
of the gas surface density, $\Sigma_g$, the radial velocity, $v_R$, and the 
ratio of the circular velocity at time $t$ to its initial value, $\frac{v_c(t)}{v_c(0)}$
starting with an inviscid equilibrium solution ({\em i.e.} with $v_R=0$ initially) and a 
nearly uniform mass density distribution.
In figures~\ref{gasprof_early} and~\ref{gasprof} we show a similar time evolution 
of these quantities but in this case we start from an initial $v_R$ consistent 
with the Navier-Stokes solution corresponding to the initial density profile we choose~\footnote{
these solutions are given in appendix A}. 
Overplotted are the analytic predictions.  
These predictions are obtained by solving the Navier-Stokes equations 
for a compressible viscous fluid in cylindrical coordinates along with the equation 
of continuity:
\begin{eqnarray}
\label{firstNS}
 \Sigma_g \left( v_R \frac{dv_R}{dr} - \frac{v_C^2}{r} \right) & = & 
- \frac{dp}{dr} -\frac{d\Phi}{dr} \nonumber \\
&+& \left(\xi + \frac{4}{3} \eta \right) 
\left( \frac{d^2 v_R}{dr^2} + \frac{1}{r} \frac{dv_R}{dr} 
- \frac{v_R}{r^2} \right) \nonumber \\ &+& 2 \frac{d\eta}{dr} \frac{dv_R}{dr} \nonumber \\ 
& + & \left(\frac{d\xi}{dr} - \frac{2}{3} 
\frac{d\eta}{dr} \right) \left( \frac{dv_R}{dr} + \frac{v_R}{r}\right)\\
\label{secondNS}
 \Sigma_g v_R \left( \frac{dv_C}{dr} + \frac{v_C}{r} \right) & = &  
\eta \left( \frac{d^2 v_C}{dr^2} +\frac{1}{r} \frac{dv_C}{dr} 
- \frac{v_C}{r^2} \right) \nonumber \\
& + & \frac{d\eta}{dr} \left( \frac{dv_C}{dr} - \frac{v_C}{r} \right)\\
\label{continuity}
 \Sigma_g \left( \frac{dv_R}{dr} + \frac{v_R}{r} \right) & = & 
- \frac{d\Sigma_g}{dr} v_R  
\end{eqnarray}
The fundamental assumption that permits the analytic calculation is that the 
circular velocity $v_C$ remains unchanged, and therefore one only needs to 
use the second Navier-Stokes equation (eq.~\ref{secondNS}) and the continuity equation 
(eq.~\ref{continuity}). However, this assumption can be tested a posteriori, when the solutions 
for density $\Sigma_g$ and radial velocity $v_R$ have been computed.
If one plugs these solutions in the first Navier-Stokes equation (eq.~\ref{firstNS}), it is easy to check
that, to second order in viscosity it is satisfied.
The analytic steady state solution of the system therefore is:
\begin{eqnarray}
\label{vcirc}
v_C(r) & = & v_0 \frac{r}{(r_0^2+r^2)^{1/2}} \\
\label{vrad}
v_R(r) & = & - \nu \frac{r}{r_0^2+r^2}  \\
\Sigma_g(r) & = & \Sigma_g(0) \frac{r_0^2 + r^2}{r^2}
\end{eqnarray} 

One important remark is that this set of solutions depends {\em only} on the shape of $v_C$
through the parameter $r_0$. Therefore, provided that the gravitational potential is fixed,
the radial inflow of gas ($v_R$ is always negative) after a certain time should be the same 
{\em whatever the initial conditions} we choose for the gas. Unfortunately, things are more 
complicated for the density profile
because the analytic solution for $\Sigma_g(r)$ diverges for $r=0$. However, there is an obvious
comment that we can make: the steady state density profile is {\em not} exponential, which implies
that it is the interplay between viscosity and star formation that must be responsible for 
turning stellar profiles into exponentials in the viscous scenario. 

The first thing to notice in figures~\ref{gasprof_early_zero},~\ref{gasprof_zero},~\ref{gasprof_early} and~\ref{gasprof} is that the BGK scheme drives the gas towards a radial
velocity profile which is in excellent agreement with the predicted
analytic profile, independent of the initial profile for $v_R$. We will show in 
section~\ref{text_diffrot} that this also holds when we start from a different initial density 
profile and/or a different gravitational potential profile, again in perfect agreement with the analytic expectations.
The situation for the density profile is different because to 
obtain a good match to the analytic fit for a density profile which 
diverges like $r^{-2}$ in $r=0$, one requires an
infinite amount of time and mass. As we can only run the computation
for a finite amount of time and handle finite numbers, we obviously cannot achieve the steady
state density distribution, even though one clearly sees that the gas will perpetually pile up 
in the center of the simulated disk. 
We also point out that the circular velocity of the gas, $v_C$, remains unchanged up to a tenth of a percent
throughout almost all of the disk, which numerically confirms
the validity of our analytic assumption that it is constant in time.
Finally, although we do not show the results in this paper,
we have also verified that decreasing $Re_c$, which is equivalent to increasing $\nu$,
results in the gas showing a similar radial flow in shape, but with a higher normalisation.

The success of BGK in giving viscous radial 
flows on the order of 1 km/s in a disk rotating differentially at 
220 km/s is remarkable. It is a technical success which is
necessary for studying star formation on a viscous time scale in disks
since chemical evolution models (Lacey and Fall 1985, Clarke 1989) find that radial
flows on that order can explain the radial distribution
of metals in disk galaxies. 

\section{Star formation in  viscous galaxy disks}
\label{SF_in_visc_disks}
\subsection{Initialization and Boundary Conditions}
\label{initial_bc}
For our simulations with star formation, the initialization is
similar to the initialization of the purely gaseous disk described in
section~\ref{visc_pure_gas}.  Here we elaborate on our initial
and boundary conditions, and discuss our motivation for choosing
this problem set-up. We initialize our gas to be in a fixed dark matter
halo and we model the dark matter halo with the logarithmic potential
defined by equation~\ref{eq:logpotential}. We choose this form for $\Phi$
because of its simplicity which allows us to perform 
analytical computations for comparison.
Furthermore this potential has the property that the rotation curve corresponding to it becomes 
asymptotically flat with velocity $v_0$ at large radii
as is observed for many spiral galaxies.  All the results presented in this
paper are from runs with $v_0$ = 220 km/s.  The total mass of the dark halo 
is chosen to be $10^{12} {\rm M}_\odot$ and that of the gaseous disk
$5\%$ of the dark matter, {\em i.e.} 
$5 \times 10^{10} {\rm M}_\odot$. Given that the gas is a small fraction of 
the dark matter, we neglect its self-gravity and assume that it evolves in 
the fixed potential created by the dark matter.

We perform our simulations with an isothermal equation of state,
meaning that we assume that at the end of each iteration
the gas instantanesouly cools to $\approx 10^4  {\rm K} $. 
We choose this temperature because it is the lowest achievable
from HI line cooling alone.  Below that temperature we would have
to worry about the formation of molecular hydrogen, which we neglect
because we do not have the necessary resolution to follow the formation
and fragmentation of molecular clouds in the context of an entire
galaxy. Accounting for a molecular component would also require us to follow an
atomic and molecular fluid simultaneously, which would considerably increase the complexity 
and computational cost of the simulation and therefore is beyond the scope
of this paper. For a $\gamma = \frac{5}{3}$ gas a temperature of 
$\approx 10^4  {\rm K} $ corresponds to a sound speed of $\approx 10$ km/s.

Since we want to initialize the gas with self-consistent viscous initial conditions, 
we set the radial flow to be given by the profile derived
from the Navier-Stokes equations (see Appendix A), assuming centrifugal
equilibrium for a particular
initial gas density profile in a potential $\Phi$ with a core radius $r_0$ (eq.~\ref{eq:logpotential}).
Indeed, due to the observed thinness of stellar 
disks in nature (scale height $\approx$ 200 pc), and the difficulty 
of dissipating stellar motions perpendicular to the galactic plane, 
it seems reasonable to assume as an initial condition for the radial
velocity that the gas collapsing to a disk in the dark matter halo was 
in approximate centrifugal equilibrium  before it started forming stars.

We choose 20 kpc for the disk radius of our simulations
because based on observations of OB stars (Chromey 1978) and
of HII regions (Fich and Blitz 1984), the stellar disk of our 
own Galaxy seems to have an outer cutoff around this value.
Our computations are performed on a 207 by 207 evenly
spaced Cartesian grid, so that for a disk with radius
20 kpc the cell size is about 200 pc per side. Again, we emphasize
that we are not trying to resolve star formation on a small
scale. We are trying to model the first order effect of turbulence, a phenomenon which might govern
star formation on a large scale.  We run each 
simulation for about 12 Gyr. On a grid of
207 X 207 this takes approximately 50,000 iterations and requires 
about 11 hours of CPU time per processor on an Origin 2000
with four r12k processors working in parallel. 

We investigate the case where the kinematic viscosity, $\nu$ is
constant because it is a simple case for which we can compute
analytic solutions for the radial velocity exactly and compare
them to the results from our hydrodynamical simulation. 
As already described in section~\ref{visc_pure_gas}, 
to achieve a constant $\nu$, we set the collision
time $\tau = \frac{2 \lambda U L}{Re_c}$. We again take $U= 220$ km/s, 
$L=20$ kpc, ${\rm Re_c} = 750$ and $\lambda = \frac{m}{2kT}$ with 
$T \approx 10^4$K.  We underscore that our choice for $\nu$ is motivated
by our intention to model a disk with a critical Reynolds number $\approx$
750, because in this way as
discussed in section~\ref{viscous_turb}, we model the lowest order effect turbulence
has on a flow, namely to turn it into a more viscous fluid with $Re = Re_c$ .

Our boundary conditions are as follows.
Outside of a radius of $R_{\rm disk}$ on our 2D Cartesian grid,
we keep two ``rings'' of ghost cells,
each ``ring'' being one cell thick. Beyond these ghost cells we do not follow
the evolution of the gas. In this way we have carved a circular grid
out of the square cartesian grid.  At the end of each updating
time step of the code, we replace the values of the velocities
in the ghost cells with the values of the closest active ring of cells
(these for which the evolution is computed).  This means that we are 
actually performing a constant radial extrapolation to reset
the values of all the macroscopic quantities (density, momentum and energy) 
in the ghost cells. Basically, this is an implementation of ``outflow'' 
boundary conditions, where the active region dictates the behaviour of the flow
inside the boundaries. 
Nevertheless, despite this a priori free boundary condition, 
we find that the mass 
flow across the outer boundaries of the grid is under control. 
At most our boundary conditions lead to a maximal loss/gain in total
mass on the grid of $1.7 \%$ (see tables~\ref{t_param} and~\ref{rot_param}).  
This was to be expected, since analytic calculations (eq.~\ref{vrad} and 
fig.~\ref{gasprof}) show that for the critical 
Reynolds number we are using, the outer radial flow should eventually settle 
around a value smaller than $0.5$ km/s. 
In the future, it will be interesting to investigate the effect of infall onto the disk
and for this we will obviously have to alter these boundary conditions.

\subsection{The star formation timescale}
\label{text_SFtimescale}

\begin{figure}
\centerline{\psfig{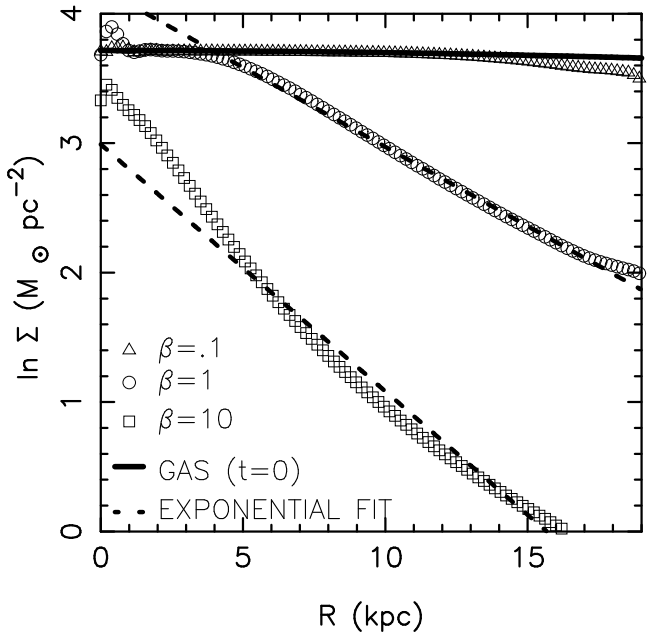}}
  \caption[]{Comparison of the final stellar profiles (after $\approx$ 12 Gyr of evolution) 
for different values of the star formation timescale to the viscous timescale, $\beta$.  
The initial radial velocity profile is derived from
the Navier-Stokes equation for the corresponding initial density profile, 
$\Sigma_g^i = \Sigma_0/( 1 + ({\frac{r}{r_p}})^{2})$ where $r_p = 78$ kpc, 
$\Sigma_0 = 41.08 {\rm M}_\odot / {\rm pc^2}$, and $r_0 = 2.56$ kpc in the gravitational potential
(see Appendix A for details).}
  \label{beta}
\end{figure}

The first point that we
wish to verify is that if star formation
proceeds on the same timescale as the viscous redistribution of
mass and angular momentum in disk galaxies then the stars
attain an exponential density profile.
We therefore define the viscous timescale, $t_{\rm visc}$,  by 
$r/{v_R}$. This estimates the time
it takes a disk annulus to move a radial distance $r$.
Our star formation law then simply reads:
\begin{equation}
\label{sfr}
\frac{d\Sigma_\star}{dt} (r,t) = \frac{\Sigma_g (r,t)}{t_\star} = \Sigma_{\rm SFR} (r,t)
\end{equation}
where, unless explicitly specified, $t_\star = t_{\rm visc}$.
In the steady state, $t_{\rm visc} = (r_{0}^{2}+r^2)/{\nu}$, so
that at large radii the assumption of a constant kinematic viscosity corresponds
to a star formation rate per unit area of $\Sigma_{\rm SFR} \propto \Sigma_g/r^2$. This
$\Sigma_{\rm SFR}$ corresponds to a Schmidt law with $n=1$ modulated by
$r^{-2}$ which means that the star formation rate at large radii is very slow.
Note that since $v_R$ varies with space {\em and} time until
the steady state is achieved, after which point it only varies with space,
$t_{\rm visc}$ (and hence $t_{\star}$) has a similar behaviour.
We assume, as do Lin \& Pringle (1987), that
once the stars form, they are frozen out of the viscous evolution
and simply move on circular orbits with a velocity given by the 
rotation curve at the location at which they form.  Hence 
we neglect any process that might give the stars a velocity
dispersion after they have formed. As a result of these assumptions, 
the implementation of star formation in the simulation only introduces 
a sink term, $-\Sigma_g (r,t)/t_\star$, in the continuity equation~\ref{continuity}.   

\begin{figure}
\centerline{\psfig{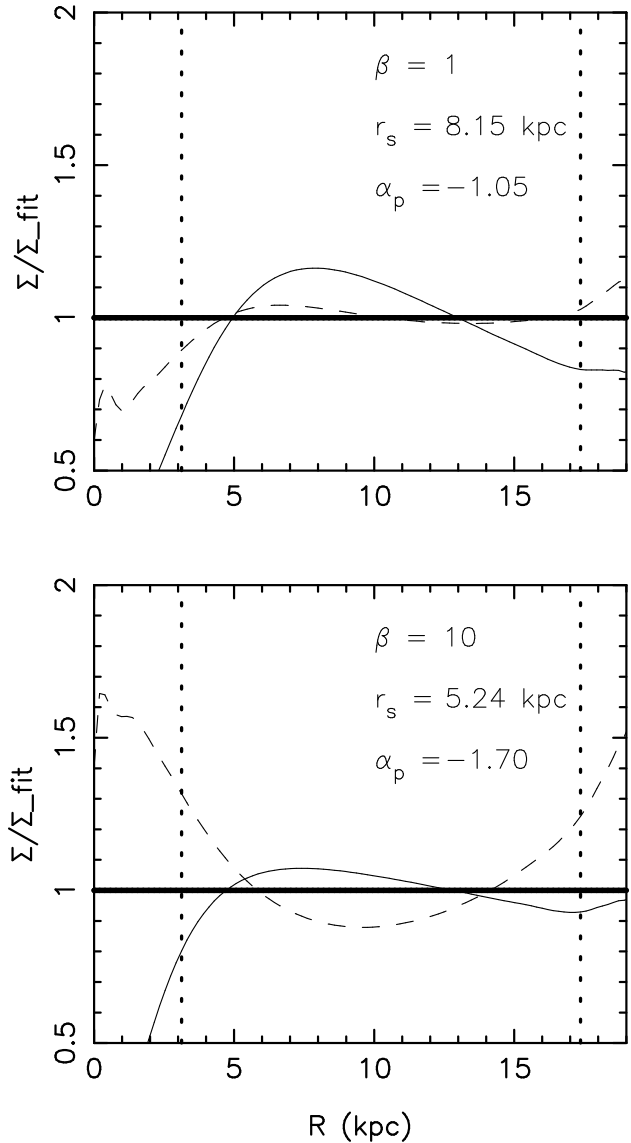}}
  \caption[]{Fitting residuals for $\beta = 1$ case (top) and $\beta = 10$ case (bottom).
Thin solid lines are the residuals for the best power law fit to the stellar profile and
the long dashed lines are the residuals for the best exponential fit. The fit is done only over
the radii enclosed within the vertical short-dashed lines. The scale length of the exponential
fit, $r_s$, and the exponent of the power-law fit, $\alpha_p$, are given on each panel.}
  \label{betares}
\end{figure}

\begin{figure*}
\centerline{\psfig{figure=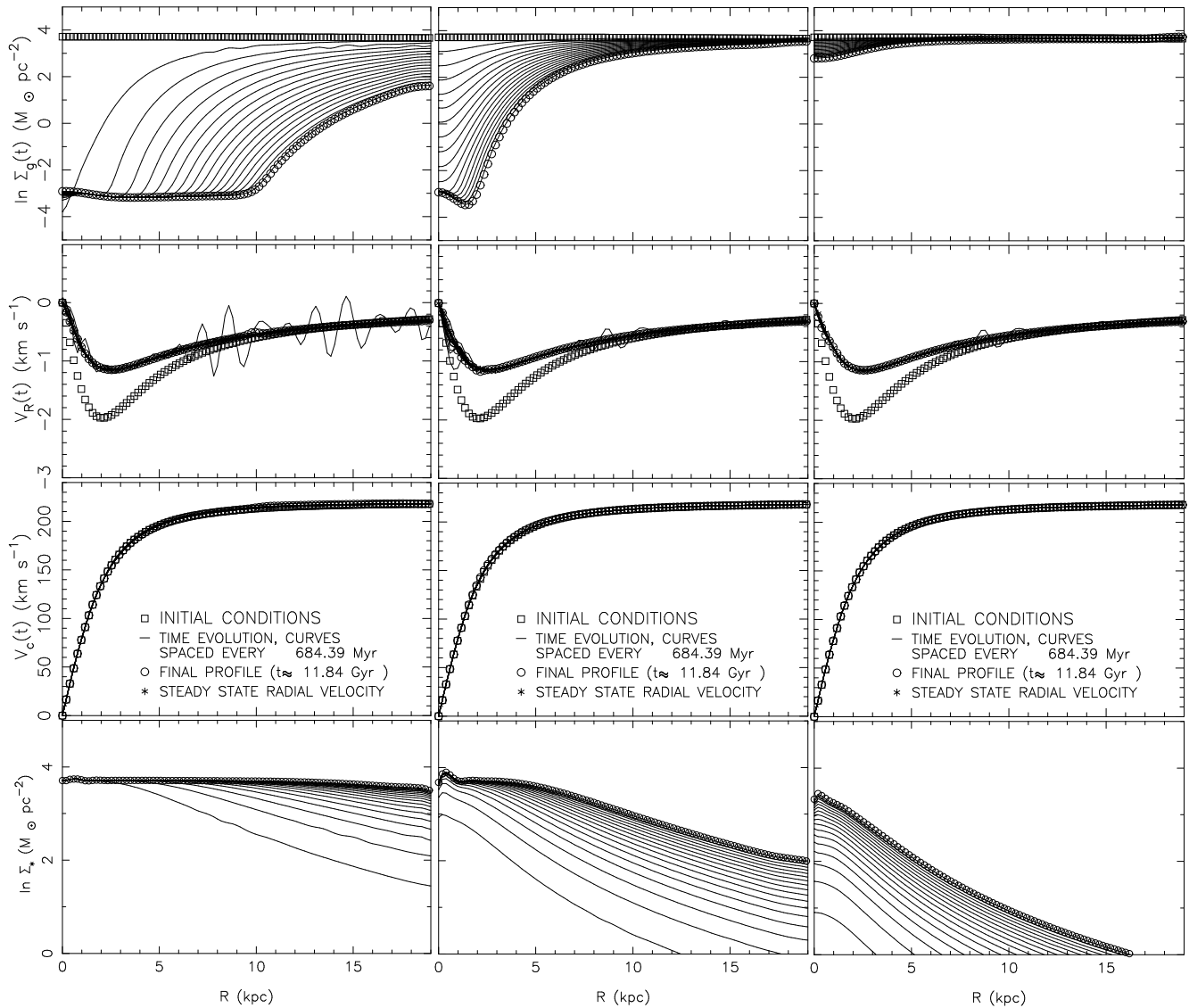,width=\hsize}}
  \caption[]{Time evolution of gas density (uppermost panels), radial velocity (second panels from top), 
tangential velocity (third panels from top)
and stellar density (bottom panels) for $\beta$ = .1 (left panels), {\em i.e.} a simulation where star 
formation takes place quickly compared to the viscous timescale, $\beta$ = 1 (middle panels), {\em i.e.}  
a simulation where star formation occurs on a
 viscous timescale and $\beta$ = 10 (right panels), {\em i.e.} a simulation where star formation is much slower 
than the viscous timescale.}
  \label{betamosaic}
\end{figure*}

We perform a set of three runs with varying
star formation timescales. Using Lin and Pringle's notation of
$\beta = t_\star/t_{visc}$, we run the code for the same three
cases that these authors considered: $\beta = .1, 1, 10$. 
The initial gas density is $\Sigma_g^i = \Sigma_0/( 1 + ({\frac{r}{r_p}})^{2})$ 
where $r_p = 78$ kpc, $\Sigma_0 = 41.08 {\rm M}_\odot / {\rm pc^2}$, and $r_0 = 2.56$ kpc 
in the gravitational potential.
Results are shown in figures~\ref{beta},~\ref{betares} and~\ref{betamosaic}.

Figures~\ref{beta} and \ref{betares} show a comparison of the stellar profiles we 
obtain after about 12 Gyr for these three different values of $\beta$.
Just as Lin and Pringle claimed, we find that the initial gas profile
is nearly frozen into the stellar profile for small $\beta$
because star formation is so rapid that almost no viscous redistribution of matter 
can take place before the stars form.  
Furthermore, as illustrated in the left panels of figure~\ref{betamosaic}
which show the time evolution of the gas density, radial velocity, and
stellar density (top to bottom), almost all the gas is converted
into stars by 12 Gyr (more than 98 \%).  This is inconsistent with present-day
spirals like the Milky-Way which still have a supply of gas of order 15 \%  
(Prantzos \& Aubert, 1995) to fuel ongoing star formation in their disks. 

At the other extreme is the case of
large $\beta$. Here too much viscous evolution occurs and the
resulting stellar profile is more a power law than an exponential (see bottom panel of
figure~\ref{betares}). Furthermore, the amount of stars formed during 12 Gyr is negligible
(a few \%), again in conflict with observations for Milky-Way type spirals which should 
have formed between 80 and 90 \% of their stars.

Only the $\beta$ = 1 case, \emph{i.e.} the case where
stars form on the viscous timescale, gives both 
approximately an exponential stellar profile (see residuals in top panel of figure~\ref{betares}) 
and a stellar density which is in better agreement with Milky Way values.
The stellar density is still low for this simulation with only about 40 \%
of stars formed, but this number depends sensitively on the steepness
of the initial gas profile (see table~\ref{t_param} and~\ref{rot_param}). If we use a 
reasonably steeper (and probably 
more realistic) initial profile we can bring the amount 
of stars formed in perfect agreement with 
the observations, {\em e.g.} for the case where $\Sigma_g^i = \Sigma_0/( 1 + ({\frac{r}{r_p}})^{4})$,
$r_p$ = 5.12 kpc, and $r_0$ = 1.92 or 2.56 kpc, about 80 \% of the disk's gas is turned
into stars after $\approx$ 12 Gyr (see section~\ref{text_diffrot}).
Looking at the lowest middle panel of figure~\ref{betamosaic}, we 
also see that after about 6 Gyr (9th thin solid
curve in the bottom middle panel, the first one being the lowest), the 
final exponential 
profile is more or less established in the sense that there is little evolution in its scale length thereafter:
the regularly spaced (in time) curves are much closer to one another than
they are at earlier times, and their slope is not changing noticeably.  
The  bottom panels of figure~\ref{betamosaic} lead to the general statement
that the evolution of 
the stellar profile proceeds much more quickly at early times than at later times because the gas
content is higher and the radial flows are generally faster in the beginning of the simulation. 

In their models of star formation in a viscous disk Firmani, Hernandez \& Gallagher (1996) 
also claim that there appear to be two distinct phases of star formation, the first corresponding
to a stage of significant viscous mass and angular momentum 
redistribution, and the second
corresponding to a stage where star formation has reduced
the amount of gas in the disk to the point that viscous evolution
is very slow and star formation proceeds slowly with an unchanging
radial velocity profile for the gas.  We remark that interpreting
this as two distinct phases of star formation must be done cautiously
because Firmani {\em{et al.}} (1996) initially take their gaseous
disk to be in inviscid centrifugal equilibrium, so that when they begin to
follow the disk evolution with equations of motion which include
viscosity, the gas first tries to find a (non-existent) equilibrium for the
viscous case and this causes a lot of initial mass and angular
momentum redistribution. For instance,
comparing our figures~\ref{gasprof_early_zero} and~\ref{gasprof_early}, we clearly see 
that the steady state for the radial velocity is reached much earlier when we start from
an initial radial velocity in agreement with the Navier-Stokes solution corresponding to the initial
density profile (middle panel of fig~\ref{gasprof_early}) than when we start from the inviscid 
equilibrium solution (middle panel of fig~\ref{gasprof_early_zero}).  
We also note that, whatever the initial conditions, the time taken to reach the steady state 
is quite short ($\approx$ 2 Gyr for the longest initial inviscid equilibrium case) compared to the 
duration of the whole simulation (about 12 Gyr).

\begin{figure*}
\centerline{\psfig{figure=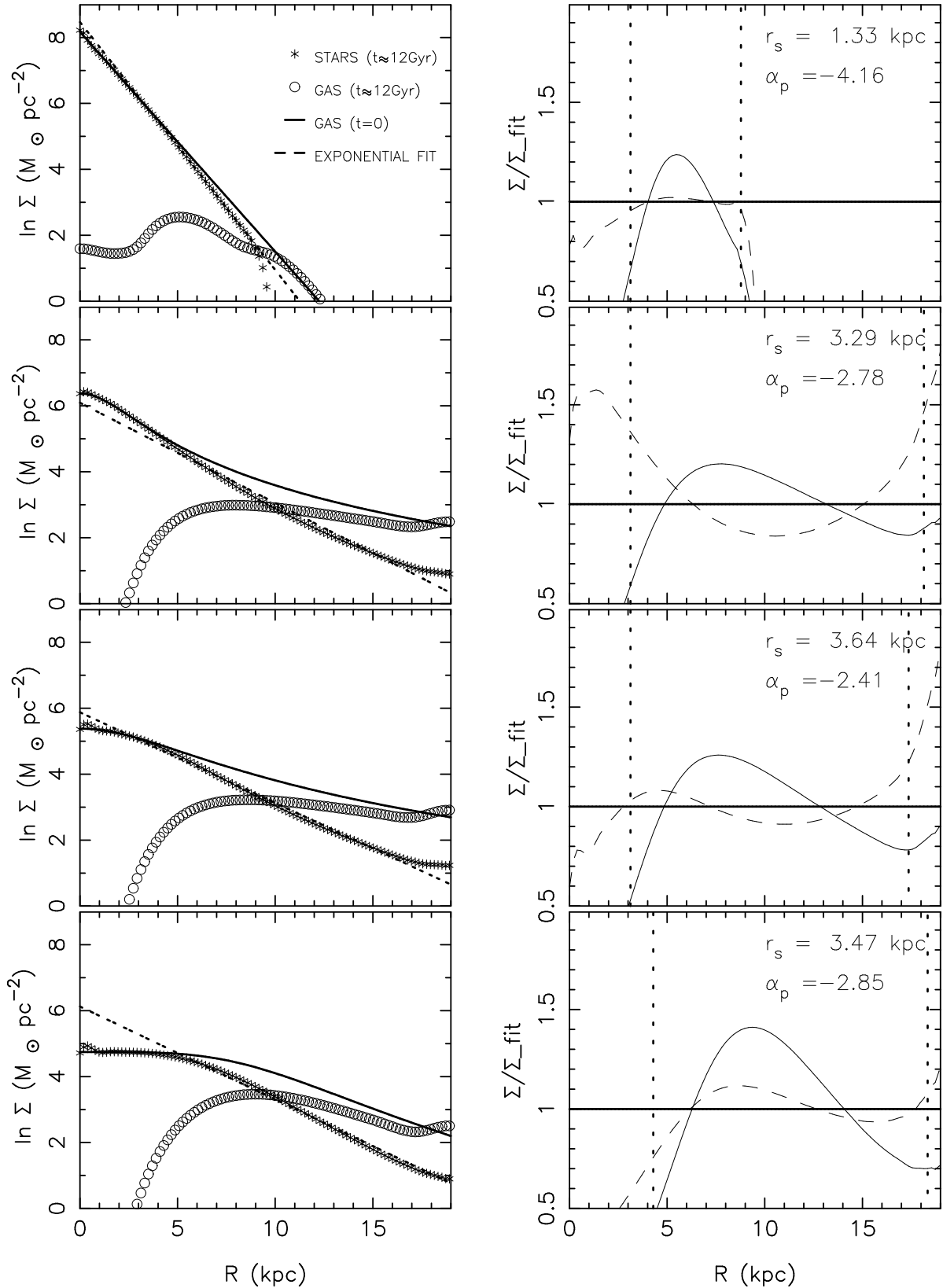,width=0.9\hsize}}
  \caption[]{Comparison of stellar and gas profiles after 12 Gyr for
varying initial gas profiles.  From top to bottom the initial gas profile
is $\Sigma_g^i = \Sigma_0 e^{-\frac{r}{r_e}}$, $\Sigma_g^i = \Sigma_0/( 1 + ({\frac{r}{r_p}})^{2})$,
$\Sigma_g^i = \Sigma_0/( 1 + ({\frac{r}{r_p}})^{2})$, and  $\Sigma_g^i = \Sigma_0/( 1 + ({\frac{r}{r_p}})^{4})$. The values of the parameters $\Sigma_0$, $r_e$, and
$r_p$ are given in table~\ref{t_param}.  The parameters defining the gravitational potential
are the same for each of these runs, namely $r_0$ = 2.56 kpc and $v_0$ = 220 km/s.
}
  \label{diffICs_fit}
\end{figure*}

Since there is yet no absolutely convincing explanation for an equality between
the viscous and star formation timescales, it is worthwhile
to explore how much leeway there is in the equivalence.
Saio \& Yoshii (1990) explored the cases where $\beta$ is .5 and 2, 
as opposed to an order of magnitude greater or smaller than $\beta = $1.
They find the $\beta = .5$ case to yield a profile close to
the pure exponential disk (van der Kruit 1987) and the $\beta = 2$
case to yield a disk with a pronounced bulge and disk. They concluded 
that the star formation timescale does not have to exactly equal
the viscous timescale to produce an exponential stellar profile.
Although we do not show results in this paper, we have run simulations with 
different $\beta$ and confirm that provided it does not differ from unity by
more than a factor 2, we get exponential profiles of a similar quality as 
those presented here. 

\begin{table*}
\begin{tabular}{|l|c|c|c|c|c|c|c|c|c|c|c|c|}
\hline
\hline
&\multicolumn{3}{c|}{Initial Parameters}&\\ 
\cline{2-4}
& & & & & & & \\ 
\multicolumn{1}{|c|}{Profile} & $\Sigma_0$ & $r_p$ & $r_e$ & $r_s$ & $r_H$ & PL & stars & gas 
& mass & $\Sigma_*$ at $R_\odot$ & $\Sigma_g$ at $R_\odot$  & $\Sigma_g$ at $R_\odot$ \\
& (${\rm M}_\odot / {\rm pc^2}$) & (kpc) & (kpc) & (kpc) & (kpc) & fit & & & loss 
& (${\rm M}_\odot / {\rm pc^2}$) & (${\rm M}_\odot / {\rm pc^2}$) & (${\rm M}_\odot / {\rm pc^2}$) \\
\hline
 & & & & & & & \\ 
PL, $\alpha_p=2$     & 41.1 & 78. & -- &  8.15 & 13.3 & -1.05 & 41.4 & 59.6 & +1.05 & 23.8 & 16.8 & 40.6 \\
EXP               & 3537. & -- & 1.5 & 1.33 & 2.52 & -4.16 & 94.3 & 5.7 & -.004 & 7.5 & 5.0 & 11.6 \\
PL, $\alpha_p=2$     & 586.9 & 2.56 & -- & 3.29 & 6.62 & -2.78 & 69.5 & 31.4 & +.84 & 28.4 & 19.6 & 40.6\\
PL, $\alpha_p=2$     & 217.3 & 5.12 & -- & 3.64 & 8.6 & -2.41 & 60.8 & 41.0 & +1.7 & 31.9 & 25.1 & 57.1\\
PL, $\alpha_p=4$     & 115.2 & 10.2 & -- & 3.47 & 5.06 & -2.85 & 59.6 & 41.5 & +1.1 & 45.2 & 31.8 & 77.2\\
 & & & & & & & \\ 
\hline
\hline
\end{tabular}
\caption []{Results from runs all having the same rotation curve ($v_0 = 220$ km/s  and $r_0=2.56$ kpc) but different initial gas profiles (PL stands for power law with exponent $\alpha_p$, and EXP stands for exponential.
See text for more details). Column \'PL fit\' contains exponents from the 
best power law fit to the final stellar profile; the following column gives the percentage of initial gas that 
has been turned into stars after 12 Gyr. From left to right, the next columns give values for the percentage
of gas left after 12 Gyr, the percentage of mass lost (or gained) from the grid after 12 Gyr, the
stellar density at the solar radius after 12 Gyr, the final gas density at the solar radius, 
and the initial gas density at the solar radius.}
\label{t_param}
\end{table*}

\subsection{Dependence on initial gas profile}
\label{text_diffICs}

The next step is to check the robustness of the $t_\star \sim t_{visc}$ hypothesis
to varying initial conditions. With this intention, we perform 
numerical experiments,
keeping the rotation curve fixed ($v_0 = 220$ km/s and $r_0=2.56$ kpc)
while varying the initial gas profile.
We have tried the following two
profiles: $\Sigma_g^i(r) = \Sigma_0 e^{-\frac{r}{r_e}}$ and
$\Sigma_g^i(r) = \Sigma_0 / ( 1 + ({\frac{r}{r_p}})^{\alpha_p})$ where
$\Sigma_0$ is a constant, $r_e$ and $r_p$ are the scale lengths
of the exponential and power law profile respectively and
$\alpha_p$ is either $2$ or $4$.  For each of the runs with rotation
curve prescribed by $v_0 = 220$ km/s and $r_0=2.56$ kpc,
Table ~\ref{t_param} presents the values for $\Sigma_0$, $r_p$, $r_e$ 
the stellar exponential scale length, $r_s$, resulting from
an exponential fit to the stars after about 12 Gyr, the initial
half mass radius, $r_H$, ({\em i.e.} the radius containing half of the mass
in the disk) the exponent from a power law (PL) fit to
the stellar profile, the percentage of initial gas that has been turned into stars after
12 Gyr, the percentage of initial gas that is left after 12 Gyr,
the percentage of initial gas mass that has been lost from the grid or gained
 after 12 Gyr,
the stellar density and the final gas density 
at the solar galactocentric radius
(8.5 kpc, e.g. Binney \& Tremaine, 1987) after 12 Gyr, and finally the initial gas density at 8.5kpc
for comparison to the sum of the stellar and gaseous densities at
that location after 12 Gyr.

We discussed the run corresponding to the first entry of
the table~\ref{t_param} in section~\ref{text_SFtimescale}.
Figure~\ref{diffICs_fit} shows a comparison of the stellar and gas profiles
for the other 4 runs presented in this table. We have overplotted the 
exponential fit 
to the final stellar profiles. We confirm that if the distribution
of gas in a disk is originally exponential, and star formation proceeds
on a viscous timescale, the exponential profile will be transfered to the stars.
However due to viscous flows, the exponential stellar
profile will have a different scale length than the scale length of
the original gaseous disk. The final exponential scale length is about
$.5$ $r_H$ which is consistent with the expectation of Lin and Pringle's (1987)
model.

For stellar profiles which did not originate from a gaseous exponential
profile, it is clear that the divergence from
the exponential is large in the central regions where all the gas has been consumed
rapidly, resulting in the initial gas profile being almost frozen into the final
stellar profile.
The central regions of real galaxies are also places where we certainly cannot neglect 
self-gravity and the presence 
of bulges and/or bars, which will have a dramatic impact on the gas.
Therefore we exclude
the inner 3-4 kpc from the fits of our exponential profiles. Additionally, 
to be conservative and
minimize a possible contamination from the boundaries, we also exclude at least the outermost
2 kpc from our fits.
The residuals, plotted in the right panel of Figure~\ref{diffICs_fit} clearly argue in favor 
of an exponential stellar profile over a power law one in most cases, although the departures from 
an exponential are not negligible and can easily reach 20 \% and more in some parts of the disk. 
Nevertheless, the profiles are generally exponential to within 10\% over several scale lengths.
Since our simulation grid only extends to 20 kpc, of which we only take a range spanning
about 14 kpc seriously, it means that for stellar profiles with a scale length of around
3.5 kpc we only have a grid big enough to tell if the exponential fit holds over 4 scale lengths,
which is not a significant improvement over Gunn's (1982) exponential-like profile. However for cases
where $r_s \approx$ 2 kpc, we obtain exponentials which hold over more than 5 scale lengths 
(see table~\ref{rot_param} and fig.~\ref{stars_rotcurve}).
 
\begin{figure}
\centerline{\psfig{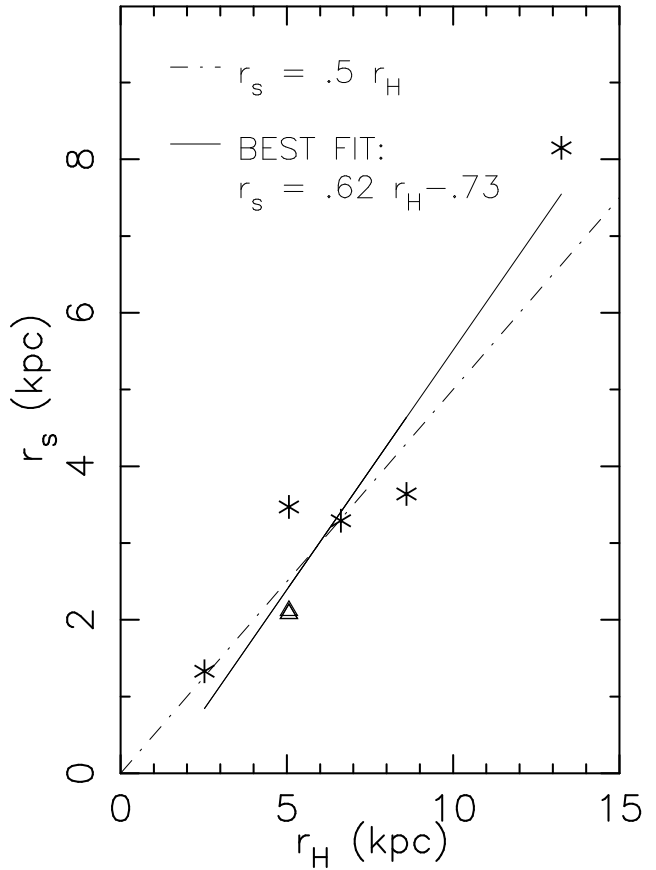}}
  \caption[]{The scale length of the stellar profile after $\approx$ 12 Gyr 
is plotted against the
initial half mass radius, $r_H$, of the gas profile.
Asterices correspond to runs with core radius $r_0 = 2.56$ kpc for the gravitational potential 
$\Phi$, but different initial gas profiles, and open triangles correspond to runs
with the same initial gas profile (cf. section~\ref{text_diffrot}) but a range of $r_0$ bracketted
by $1.92$ and $5.12$ kpc.}
  \label{halfrm_rs}
\end{figure}

On another note, in agreement with Lin \& Pringle (1987) we find that the final stellar
exponential scale length, $r_s$, can be roughly approximated by half the
initial half-mass radius, $r_H$, for the gas (fig.~\ref{halfrm_rs}).  
Hence, even though the final stellar profile is roughly exponential 
and in its shape forgets the initial gas profile, the magnitude of the
stellar exponential scale length preserves the memory of the
degree of central concentration of the initial gas distribution.

After 12 Gyr we find that there is about $6 - 60 \%$ gas left
in the galaxy depending on the steepness of the initial gas profile (see tables~\ref{t_param} and 
\ref{rot_param}), and obtain agreement with the final gas fraction of present day disks ($\approx$ 15 \%) only for quite steep profiles. 
This is of considerable importance, because it means that by simply equating the 
star formation timescale 
to the viscous timescale, for disks with initially steep
profiles we not only recover the exponential stellar profile, 
but we naturally get the star formation efficiency correct. We also find that
our stellar densities at the solar galactocentric radius, $R_{\odot}$ = 8.5 kpc
are on the low side of the Milky Way value ($35\pm5$${\rm M}_\odot / {\rm pc^2}$, Kuijken and Gilmore 1989)
for steep profiles, but not completely incompatible with it, provided a moderate amount of 
infall takes place during the 12 Gyr of evolution to boost the star formation rate.
To our knowledge, ours is the first investigation 
that does not evolve the viscous disk in arbitrary time units, and then
associate the present time with the time corresponding to the case
where the stellar profile has solar neighborhood values at the solar radius
(see for example Olivier {\em et al.} 1991, Ferguson and Clarke 2001). 
We stress that it is mandatory to have real physical units to be able to comment, as we do above, 
on whether the viscous timescale is a natural timescale for star formation.

\subsection{Dependence on rotation curve}
\label{text_diffrot}

\begin{figure*}
\centerline{\psfig{figure=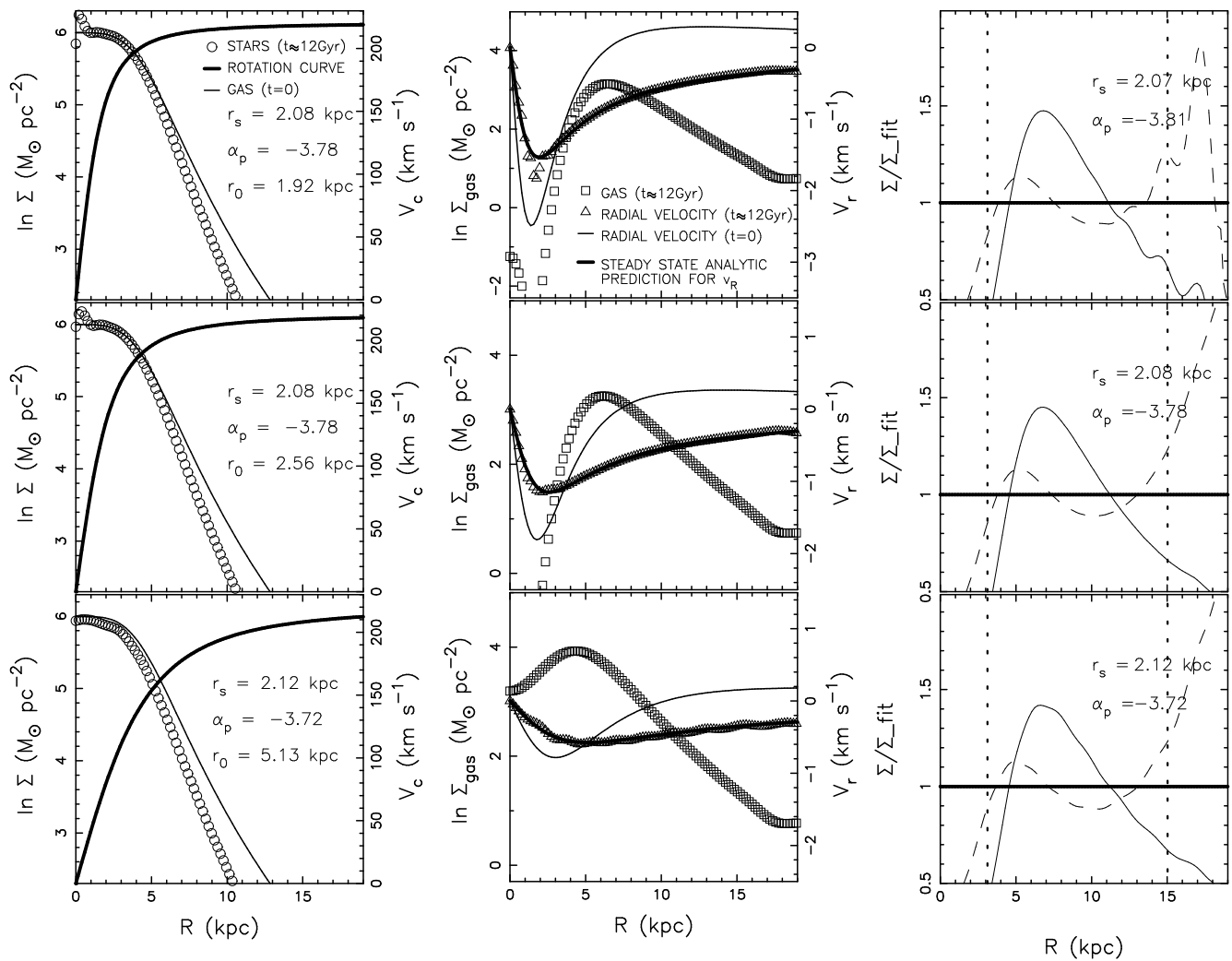,width=\hsize}}
  \caption[]
{Comparison of the stellar (left), gas and radial velocity (middle) profiles 
after $\approx$ 12 Gyr for different rotation curves obtained by varying the core radius $r_0$ of
the gravitational potential $\Phi$ (from
top to bottom $r_0 = $ 1.92, 2.56, 5.12 kpc). The initial gas profile
for each of these runs is $\Sigma_g^i = \Sigma_0/( 1 + (\frac{r}{r_p})^4)$
with $r_p = 5.12$ kpc and $\Sigma_0=402 {\rm M}_\odot / {\rm pc^2}$. The residuals
for the best exponential (dashed line) and the best power law (thin solid line) fits, within
the radii delineated by the vertical short dashed lines, are given in the right hand
panel.}
  \label{stars_rotcurve}
\end{figure*}

Lin and Pringle (1987) also stipulate that
a stellar exponential profile always arises if $t_\star \approx t_{visc}$
regardless of the shape of the dark matter potential.
By varying ${\rm r_0}$ in the expression for the dark halo
potential, $\Phi$, we  explore star formation for viscously evolving disks
in potentials ranging from those whose rotation curves rise steeply in the central regions to those with
more slowly rising curves.  For each of the runs for which
we present results, the initial gas density profile is 
$\Sigma_g^i = \Sigma_0/( 1 + (\frac{r}{r_p})^4)$
with $r_p = 5.12$ kpc and $\Sigma_0=402 {\rm M}_\odot / {\rm pc^2}$.
We again start from the initial radial velocity 
profile consistently derived from the Navier-Stokes solution 
(see Appendix A).  

Figure~\ref{stars_rotcurve} shows a comparison of the stellar profiles after
about 12 Gyr for the cases where $r_0 = 1.92, 2.56, 5.12$ kpc.
This figure also shows the corresponding radial velocity and gas profiles
at about 12 Gyr. As mentioned in section~\ref{visc_pure_gas}, we emphasize that the steady state radial 
velocity profile is now different for these three simulations, because it depends on $r_0$.
Once again, we draw the attention of the reader to the fact that the BGK scheme converges 
to the analytic solution for the radial velocity (even if less rapidly in the case of the most slowly rising rotation curve). 
Like Lin \& Pringle (1987) who state that the results of viscous evolution are insensitive to the 
central concentration of the dark matter, we find that the stellar 
densities reach only slightly higher values and have slightly steeper scale lengths ($r_s = 2.07$ kpc for 
$r_0 = 1.92$ kpc and 2.12 kpc for $r_0 = 5.12$ kpc)
when the rotation curve rises more steeply.  Once again, we are aware that for the
high central densities we have here, self-gravity becomes important and we plan
to tackle this issue in a subsequent paper. In this work we simply exclude 
the central region of the disk from the fit. 
\begin{table*}
\begin{tabular}{|l|c|c|c|c|c|c|c|c|c|c|c|c|}
\hline
\hline
&\multicolumn{3}{c|}{Initial Parameters}&\\ 
\cline{2-4}
& & & & & & & \\ 
\multicolumn{1}{|c|}{Profile} & $\Sigma_0$ & $r_0$ & $r_e$ & $r_s$ & $r_H$ & PL & stars & gas 
& mass & $\Sigma_*$ at $R_\odot$ & $\Sigma_g$ at $R_\odot$ & $\Sigma_g$ at $R_\odot$ \\
& (${\rm M}_\odot / {\rm pc^2}$) & (kpc) & (kpc) & (kpc) & (kpc) & fit & & & loss 
& (${\rm M}_\odot / {\rm pc^2}$) & (${\rm M}_\odot / {\rm pc^2}$) & (${\rm M}_\odot / {\rm pc^2}$) \\
\hline
 & & & & & & & \\
PL, $\alpha_p=4$   & 402. & 1.92 & -- & 2.08 & 5.07 & -3.78 & 82.4 & 17.6 & -.02 & 28.2 & 17.4 & 45.4 \\
PL, $\alpha_p=4$   & 402. & 2.56 & -- & 2.08 & 5.07 & -3.78 & 81.3 & 18.7 & -.01 & 27.6 & 17.9 & 45.4 \\
PL, $\alpha_p=4$   & 402. & 5.12 & -- & 2.12 & 5.07 & -3.72 & 71.8 & 28.2 & +.01 & 24.4 & 21.0 & 45.4 \\
 & & & & & & \\ 
\hline
\hline
\end{tabular}
\caption[]{Results from runs starting with the same initial gas profile 
($\Sigma_g^i = \Sigma_0/(1 + (\frac{r}{r_p})^{4})$, where $r_p=5.12$ kpc) but different rotation curves 
obtained by changing the value of $r_0$. $v_0$, the asymptotic velocity is 220 km/s for all these runs. 
Columns are the same as those of table~\ref{t_param} except $r_p$ (third column) has been replaced
by $r_0$.}
\label{rot_param}
\end{table*}

Table~\ref{rot_param}
summarizes the findings of these runs. Even though now not all the rotation
curves are good representations of the Milky Way rotation curve, we
take the solar galactocentric radius value of density as a rough estimate of
what we should find.  We again see that we match
the stellar surface densities in the solar neighborhood reasonably
well and the quantity of gas left over in the disk after $\approx 12$ Gyr
is consistent with the observational estimate of $\approx$ 15 \%. Furthemore, the
relationship between the stellar exponential scale length
and the initial half mass radius of the gas, ($r_s$ = .5 $r_H$) still
holds (see triangles in fig.~\ref{halfrm_rs}).  
We also learn from figure~\ref{halfrm_rs} that for a fixed initial gas profile,
the steepness of the rotation curve (quantified by $r_0$ in $\Phi$)
introduces very little spread in the correlation between initial gas half mass
radius and the stellar exponential scale length after 12 Gyr.  

Another thing we can gauge from these runs, is how large an effect the magnitude
of the radial velocity has on the final stellar density.
Fig.~\ref{stars_rotcurve} and table~\ref{rot_param} clearly show  that the 
magnitude of the radial velocity
is correlated to the total amount of stars formed at 12 Gyr.  For example
the run with the most slowly rising rotation curve ($r_0 = 5.12$ kpc) has a lower 
steady-state radial velocity throughout the disk than the runs with steeper rotation curves. 
In the steady state, the maximum value of $v_R$ for the run with $r_0 = 5.12$ kpc 
is a factor 1.8 smaller than the value for the $r_0 = 2.56$ kpc run. 
Since $t_\star \sim t_{visc}$, where $t_{visc} = r/v_R$, the star formation timescale
is longer for the run with $r_0 = 5.12$ kpc and hence at 12 Gyr
it has formed 10 \% fewer stars than the run with $r_0 = 2.56$ kpc.  
This points to another important consideration,
namely the value we take for the critical Reynold's number, $Re_c$, which determines
the amount of viscosity and hence the magnitude of the radial flow. We took
$Re_c \sim 750$, but as we discussed in section~\ref{viscous_turb}
it could range from $10 - 10^3$. If we had taken $Re_c \sim 100$, for example,
then the collision time in the scheme, and hence the kinematic viscosity would
increase by a factor 7.5 causing the radial inflow velocities to have a maximum
velocity of about 7.5 km/s. Given the lack of observations which pinpoint the radial
velocity of gas in slowly evolving disks, we rely on arguments concerning 
radial metallicity distributions (Lacey and Fall (1985), Clarke (1989))
to constrain the magnitude of the radial velocity to about 1 km/s,
but we underline that this is indirect evidence for the magnitude of 
$v_R$ and hence it is still debatable.

Finally we remark on the gas profiles. 
Consistent with Lacey and Fall's (1985) results, we find that the gas surface density
profiles at about 12 Gyr are deficient in gas at radii less than about 5 kpc
(Figs.~\ref{diffICs_fit} and ~\ref{stars_rotcurve}).
Also in general we find that the gas profiles after 12 Gyr for radii greater
than 5 kpc, if fit by an exponential,
have a larger scale length than the stellar profile. This is also
consistent with Lacey and Fall's results. Nevertheless, we do not want to overinterpret
our results for the gas since the gas profiles can be strongly influenced both by
feedback effects induced by supernovae or stellar winds and infall.

\subsection{Star Formation Rates}
\label{text_starformationrates}

\begin{figure}
\centerline{\psfig{figure=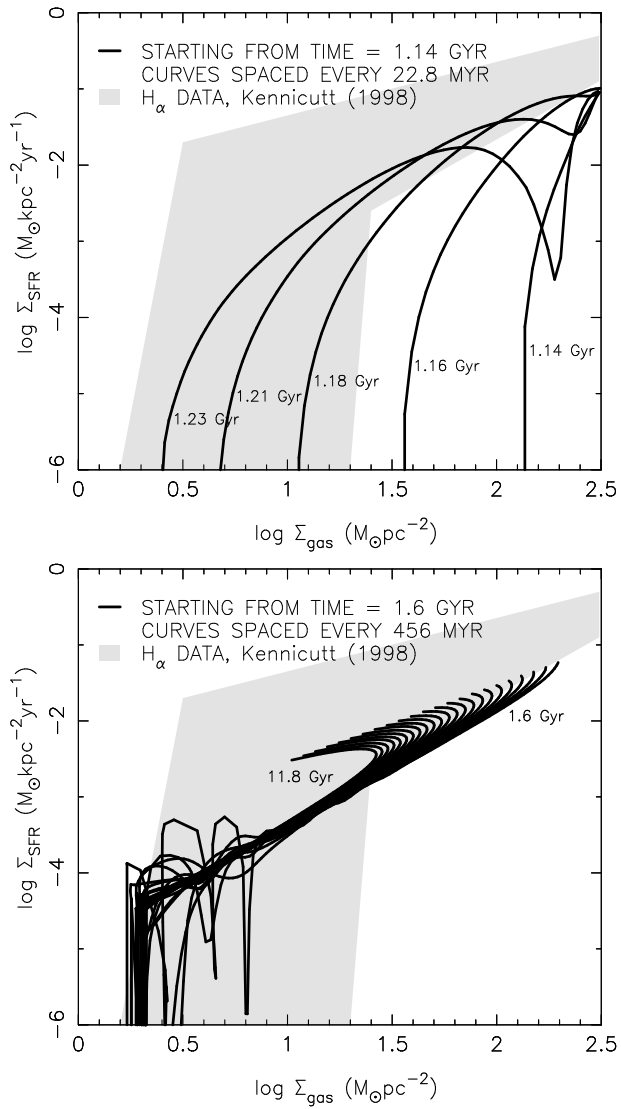,width=\hsize}}
  \caption[]{Time evolution of the star formation rate surface density as a function 
of the gas surface density in one of our simulations defined by 
$\Sigma_g^i = \Sigma_0/(1 + (\frac{r}{r_p})^{4})$, with $r_p=5.12$ kpc, $r_0 = 2.56$ kpc, 
and $v_0 = 220$ km/s. Solid lines are our star formation 
rate profiles at times indicated on the figure. Shaded region is observational data by Kennicutt (1998).}
  \label{logsfr_sigma}
\end{figure}

Another test of the viscous model for star formation in galactic disks,
is the comparison of the star formation rate with the gas
surface density as a function of time.  

On figure~\ref{logsfr_sigma}, we plot the results from our model with
an initial gas profile given by $\Sigma_g^i = \Sigma_0/(1 + (\frac{r}{r_p})^{4})$, 
with $r_p=5.12$ kpc, and evolving in a potential 
characterized by $r_0 = 2.56$ kpc and $v_0 = 220$ km/s (other results for
this profile were shown in fig.~\ref{stars_rotcurve}).
On the same plot we also indicate the observational constraints
set by Kennicutt's H$\alpha$ data (Kennicutt 1998) for the SFR surface density $\Sigma_{\rm SFR}$ and
CO data for the gas surface density, $\Sigma_{gas}$. 
Top panel of fig.~\ref{logsfr_sigma} gives results of the early evolution (up to 1.2 Gyr), 
whereas bottom panel of the same figure shows the late evolution (from 1.6 to 12 Gyr).
There are two important remarks to make concerning this figure: the first is that 
one can clearly see 
star formation happening inside out, with the central region having the highest 
star formation rate in the beginning. The latter progressively decreases due to the lack of 
gas to fuel it at late times. 
But this monotonic behaviour only takes place after the radial velocity profile has settled down 
to the steady state solution, because we see in the top panel of fig.~\ref{logsfr_sigma} that there are
some regions with a high density of gas that form a low amount of stars in the early stage. This is due to 
outflows of gas for which we quench star formation. We explicitly only allow stars to form
where the radial velocity is directed towards the center of the disk.  The second point 
is that although we
naturally reproduce the slope of the relationship and the cut-off at low gas surface densities seen in the data, 
the locus occupied by our curves at late stages is offset towards low values. This, however, should not be
considered as a failure of the model, as the normalisation of the total gas mass is somewhat arbitrary.
Had we started the simulation with 15 \% of baryons in our dark matter potential
instead of 5 \% (which is also a reasonable cosmological value for the baryon fraction), the star formation 
rates and gas surface densities would have increased by a factor of three, bringing the model in perfect 
agreement with the observational data. Also, the picture would be more complicated if we introduced 
even a moderate amount of infall, because both star formation rate and gas surface density would also
go up. Furthermore, as discussed at the end of section~\ref{text_diffrot}, had we taken a higher value
for the viscosity, we would have found faster star formation timescales which would also have
increased the star formation rates.

\subsection{Angular momentum}

Finally, for the rotating viscous disk, one might worry that, as 
pointed out by Lynden-Bell and Pringle (1974), the equilibrium is 
reached when a portion of fluid with negligible mass 
carries all the angular momentum to infinity while the rest of the gas
collapses to the center. This would only exacerbate the angular 
momentum problem seen in N-Body + SPH simulations, if it were not for 
the fact that star formation has to proceed on a viscous timescale 
in order for galaxies to end up with exponential light profiles, so the redistribution of 
angular momentum cannot be very efficient.  Although 
in the simulations presented in this paper, we do not take into account feedback 
via supernovae explosions or stellar winds that would reheat and redistribute the gas, 
we show that viscous evolution does not lead to a dramatic decrease in the specific 
angular momentum of the gas.

\begin{figure*}
\centerline{\psfig{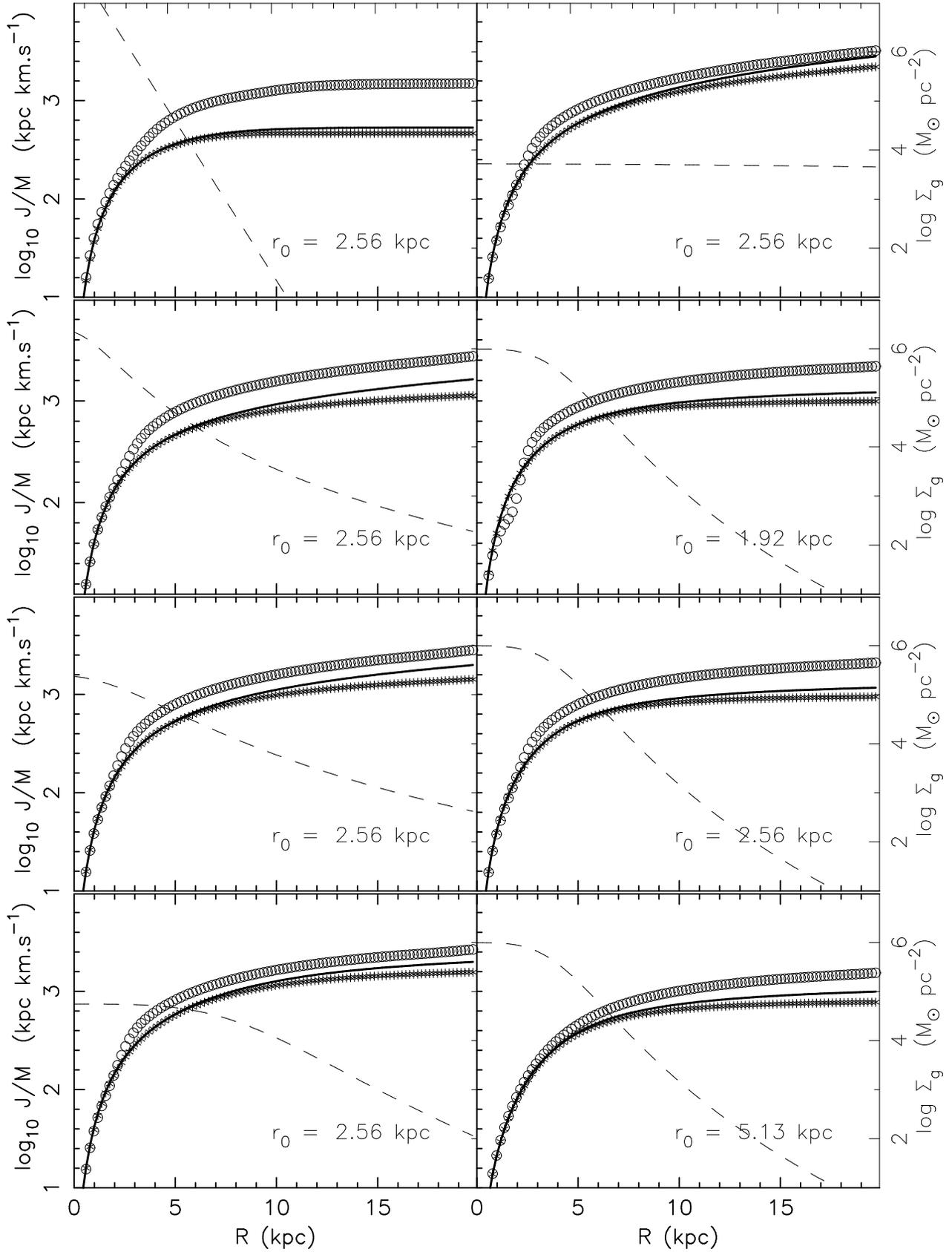}}
  \caption[]{Comparison of the cumulative specific angular 
momentum profiles for the initial gas distribution (solid line) and both the final stellar (asterices) 
and gas (open circles) distributions, {\em i.e.} after 12 Gyr of viscous evolution.
The dashed line shows the initial gas profiles. The exact analytic shapes of these profiles are
given in tables~\ref{t_param} and ~\ref{rot_param}.  
The core radius, $r_0$,  of the gravitational potential is written in the lower right
corner of each panel and $v_0$, the asymptotic circular velocity is 220 km/s in all simulations.}
  \label{spec_angmom}
\end{figure*}

On figure~\ref{spec_angmom} we plot the time evolution of the 
cumulative specific angular momentum, $j_c = J/M$, of both gas and stars. This quantity 
is obtained by computing the total amount of angular momentum in concentric gaseous/stellar disks and dividing
it by the gaseous/stellar mass in these disks. There are a couple of things that we would like to emphasize 
on this figure. First is that in all the runs, the stellar angular momentum distribution closely follows
the initial gas angular momentum distribution, which already points out that viscous redistribution of 
angular momentum must be small. This is particularly noticeable in the inner 
regions of the disk. Second, one can immediately see from figure~\ref{spec_angmom} that the final 
gas angular momentum distribution is always higher than the initial one, especially at large radii.   
In fact, there is a clear trend that the steeper the initial gas density profile and rotation curve, 
the larger the discrepancy between the initial and final specific angular momenta of the gas. 
The reason for this is the following: since 
initial gas density profiles are generally decreasing functions of radius, equation~\ref{sfr} combined
with equation~\ref{vrad} imply that the star formation rate is much higher in the center of disk galaxies.
Hence star formation will proceed inside out, consuming gas with low specific angular momentum first. 
As a result, after 12 Gyr, the leftover gas is naturally located in the outer regions of the disk where 
the specific angular momentum is higher in general, and the viscous timescale so large that 
redistribution is minimal, naturally yielding an increase in the cumulative specific angular momentum.

\section{Conclusions}
\label{conclusions}

We have performed the first numerical hydrodynamical simulations probing 
the viscous evolution of isothermal and non self-gravitating disk galaxies embedded
in static dark halo potentials.

Despite the simplicity of such a model, where we have also neglected the return
fraction of gas through stellar evolution, infall, and feedback, 
we obtain disks with approximately exponential stellar profiles,
regardless of their initial gas profile and rotation curve 
as long as $t_\star \sim t_{visc}$.  
Moreover, we obtain fractions of gas and stars after $\sim$ 12 Gyr of viscous evolution
that are consistent with observations, without having to invoke an
efficiency parameter for star formation.  Although
a detailed explanation for the connection between the viscous timescale
and star formation is yet to be worked out in details, this strongly suggests that 
the viscous timescale might indeed be the natural timescale for star formation. 
It is also a more appealing alternative than fine-tuning
the initial conditions of proto-galactic clouds destined to form disks in order 
to obtain exponential stellar density profiles.

The specific results of our investigation are:
\begin{itemize}
\item When $t_\star \sim t_{visc}$, the resulting stellar profile
is approximately exponential.  When $t_\star$ is 10 times
smaller than the viscous timescale then the initial gas profile is
frozen into the stellar distribution and when $t_\star$ is 10 times
larger than the viscous timescale then the stellar distribution
is better described by a power law.
\item Exponentials fit reasonably well the stellar profiles
after $\sim$ 12 Gyr, regardless of the initial gas profile or the
rotation curve. Furthermore, there is no requirement on conservation
of specific angular momentum throughout star formation.  
\item For rotation curve profiles similar to Milky Way profiles,
($r_0 = 2.56$ kpc and $v_0 = 220$ km/s in $\Phi$), after $\sim$ 12 Gyr the stellar densities
in the solar neighborhood  and 
the final gas fraction in the disk are consistent with observations, provided 
the initial gas density profile is quite steep.
This suggests that the viscous timescale gives the star formation
efficiency naturally.  
\item The disk scale length, $r_s$, depends most strongly on the
degree of central concentration of the initial gas density profile 
(figure~\ref{halfrm_rs}). The
steepness of the dark halo rotation curve introduces very little
spread in the initial gas half mass radius, $r_H$,  versus $r_s$ 
relation.  
\item The cumulative specific angular momentum of the final distribution of
gas is always higher than the initial cumulative specific
angular momentum of the gas due to inside out formation.  
The steeper the initial gas profile, the more important is the discrepancy
between the cumulative specific angular momentum of
the gas after $\sim$ 12 Gyr and its initial cumulative specific
angular momentum. The cumulative specific angular momentum of the stars
after $\sim$ 12 Gyr, on the other hand, closely matches the initial cumulative specific angular
momentum of the gas.
\item The slope of the star formation rate density versus gas surface density 
in the viscous disk models is in fair agreement with the observed one, and the 
cut-off seen at low gas surface densities in the observations is naturally reproduced. 
\end{itemize}

Clearly, this work is a necessary first step which proves the ability of the BGK 
scheme to follow the physical processes occuring in a galactic disk. In following works, 
we shall focus on implementing more realistic conditions for the starting point of the 
simulation and the mechanisms driving the disk evolution. We also plan to extend the code to 
3D and include self-gravity of both dark halo and gas.

\acknowledgements{We are pleased to thank Kun Xu for providing us
with a  version of the BGK scheme employing a reconstruction of
g from two half-Maxwellians. We also thank St\'ephane Colombi for
parallelizing the scheme.  AS thanks Kevin Prendergast for invaluable
discussions and Oxford University for its
hospitality during the course of much of this work. }

\appendix

\section{Initial conditions}

Here we give analytic expressions for the initial radial velocities, $v_R^i$, which correspond 
to the initial density profiles, $\Sigma_g^i$ we use in the paper, {\em i.e.} knowing
$v_C$ from equation~\ref{vcirc}, we compute $v_R^i$ by solving the second Navier-Stokes equation 
(eq.~\ref{secondNS}). This yields: 
\begin{itemize}

\item if  $\Sigma_g^i(r) =  \Sigma_0$~=~constant,
\begin{equation}
v_R^i(r) =  - \nu \frac{r(r^2+4r_0^2)}{(r_0^2+r^2)(2r_0^2+r^2)}  \\
\label{cstini}
\end{equation} 

\item if  $\Sigma_g^i(r) =  \Sigma_0 / [1+ (\frac{r}{r_p})^2]$
\begin{equation}
v_R^i(r) =  - \nu \frac{r(4r_p^2r_0^2+r^2 r_p^2+2r^2r_0^2-r^4)}{(r_0^2+r^2)(2r_0^2+r^2)(r_p^2+r^2)}  \\
\label{p2ini}
\end{equation} 

\item if  $\Sigma_g^i(r) =  \Sigma_0 / [1+ (\frac{r}{r_p})^4]$
\begin{equation}
v_R^i(r) =  - \nu \frac{r(4r_p^4r_0^2+r^2r_p^4-3r^6)}{(r_0^2+r^2)(2r_0^2+r^2)(r_p^4+r^4)}  \\
\label{p4ini}
\end{equation} 

\item if  $\Sigma_g^i(r) =  \Sigma_0 \exp(-r/r_s)$
\begin{equation}
v_R^i(r) =  - \nu \frac{r(4r_sr_0^2+r^2r_s^2-rr_0^2-r^3)}{(r_0^2+r^2)(2r_0^2+r^2)r_s}  \\
\label{expini}
\end{equation} 

\end{itemize}

Note that these initial conditions for the radial velocity have quite different asymptotic properties.
For instance, the initial radial velocity corresponding to the constant density profile (eq~\ref{cstini}) 
is always negative, meaning the gas is constantly flowing in as the steady state radial velocity is also always 
negative (eq~\ref{vrad}). However, both radial velocities go to zero like $r^{-1}$ at large radii for the 
initial power profiles (eq~\ref{p2ini} and~\ref{p4ini}), which means that there should be a small initial outflow
of gas in these simulations. Finally the initial radial velocity profile corresponding to an initial exponential
density profile (eq~\ref{expini}) can have arbitrarily high outflows at large radii, depending on the value
of $r_s$: the smaller the value, the more important the outflow since $v_R^i \propto r_s^{-1}$.

%\clearpage
%\begin{figure*}
%%\epsfbox{gasprof.ps}
%%\epsfbox{}
%\centerline{\psfig{figure=sb_halfm.ps,width=.55\textwidth,angle=0}}
%  \caption[]{The central stellar surface density of the gas profile
%vs. the scale length of the stellar profile after $\approx$ 12 Gyr.
%Asterices correspond to runs with $r_0 = 2.56$ kpc in $\Phi$ but
%different initial gas profiles, open circles correspond to runs
%with $r_0 = 10.24$ kpc in $\Phi$ but
%different initial gas profiles, and open triangles correpond to runs
%all with a constant initial gas profile but a range of $r_0$ in $\Phi$
%from $1.28$ to $10.24$ kpc.}
%  \label{sb_halfrm}
%%%  \label{}
%\end{figure*}
%\clearpage
%\begin{figure*}
%%%\epsfbox{gasprof.ps}
%%\epsfbox{}
%\centerline{\psfig{figure=sfrprofunir200_bw.ps,width=.95\textwidth,angle=0}}
%  \caption[]{Comparison to H$\alpha$ data of the correlation between
%the star formation rate and gas density.  The results come from our model with
%a uniform initial gas profile evolving in a potential 
%characterized by $r_0 = 10.24$ kpc and $v_0 = 220$ km/s.
%}
%  \label{logsfr_sigma}
%%%  \label{}
%\end{figure*}


\begin{thebibliography}{}

\bibitem[]{} Barnes J. E. \& Efstathiou G., 1987, ApJ, 319, 575
\bibitem[]{} Bhatnagar P.L., Gross E.P., Krook M., 1954, Phys. Rev., {94}, {511}
\bibitem[]{} Binney J. \& Tremaine S., 1987, Galactic Dynamics, Princeton University Press
\bibitem[]{} Chromey F. R., 1978, AJ, 83, 162
\bibitem[]{} Clarke C. J., 1989, MNRAS, 238, 283
\bibitem[]{} Cole S., Aragon-Salamanca A., Frenk C. S., Navarro J. F. \& Zepf S. E., 1994, MNRAS, 271, 781
\bibitem[]{} Contardo G., Steinmetz M. \& Fritz-von Alvensleben U., 1998, ApJ, 507, 497
\bibitem[]{} Crampin D. J. \& Hoyle F., 1964, ApJ, 140, 99 
\bibitem[]{} Dalcanton J. J., Spergel D.N., Summers F. J., 1997, ApJ, {482}, 659
\bibitem[]{} Duschl W. J., Strittmatter P. A. \& Biermann P. L, 2000, A\&A, 357, 1123
\bibitem[]{} Ferguson A. \& Clarke C. J., 2001, MNRAS, in press, astro-ph/0103205
\bibitem[]{} Fich M. \& Blitz L., 1984, ApJ, 279, 125
\bibitem[]{} Firmani C., Hernandez X. \& Gallagher J., 1996, A\&A, 308, 403
\bibitem[]{} Gerritsen J. P. E. \& Icke V., 1997, A\&A, 325, 972  
\bibitem[]{} Gunn J. E. , 1982, in Astrophysical Cosmology, ed. H. A. Br\"uck,
G. V. Coyne, and M. S. Longair (Vatican: Pontificia Academia Scientiarum), 248
\bibitem[]{} Hernquist L., 1990, ApJ, 356, 359 
\bibitem[]{} Huynh H. T., 1995, SIAM J. Numer. Anal., 32, 1565
\bibitem[]{} Kauffmann G. A. M., White S. D. M. \& Guiderdoni B., 1993, MNRAS, 264, 201
\bibitem[]{} Katz N., 1992, ApJ, 391, 502
\bibitem[]{} Katz N. \& Gunn J. E., 1991, ApJ, 377, 365
\bibitem[]{} Kennicutt R. C., 1998, ARA\&A, 36, 189
\bibitem[]{} Kim C. \& Jameson A.,  1998, J. Comput. Phys., in press
\bibitem[]{} Kuijken K. \& Gilmore G., 1989, MNRAS, {239}, 605 
\bibitem[]{} Lacey C. G. \& Fall, M., 1985, ApJ, {290}, 154
\bibitem[]{} Lin D. C. N. \& Pringle J. E., 1987, ApJ, {320}, L87 
\bibitem[]{} Lynden-Bell D. \& Pringle J. E., 1974, MNRAS, 168, 603
\bibitem[]{} Mestel L., 1963, MNRAS, {126}, 553
\bibitem[]{} Mihos J. C. \& Hernquist L., 1994, ApJ, 437, 611
\bibitem[]{} Mo H., Mao S. \& White S. D. M., 1998, MNRAS, {295}, 319
\bibitem[]{} Navarro J. F., Frenk C. S. \& White S. D. M., 1996, ApJ, 462, 563
\bibitem[]{} Navarro J. F. \& Steinmetz M., 1997, ApJ, 478, 13
\bibitem[]{} Olivier S., Blumenthal G. R., Primack J.R.,  1991, MNRAS, {252}, 102 
\bibitem[]{} Prantzos N. \& Aubert O., 1995, A\&A, 302, 69
\bibitem[]{} Prendergast K. H., Xu K., 1993, J. Comput. Phys., {109}, 53
\bibitem[]{} Press W. H. \& Schechter P., 1974, ApJ, 187, 425
\bibitem[]{} Richard d. \& Zahn J.-P., 1999, A\&A, 347, 734
\bibitem[]{} Saio H. \& Yoshii Y., 1990, ApJ, {363}, 40
\bibitem[]{} Sanders R. H. \& Prendergast K. H., 1974, ApJ, {188},489
\bibitem[]{} Silk J.  \& Norman C., 1981, ApJ, {247}, 59
\bibitem[]{} Silk J., 2001, MNRAS, in press, astro-ph/0010624 
\bibitem[]{} Slyz A. \& Prendergast K. H., 1999, Astron. Astrophys. Suppl. Ser., {139}, 199
\bibitem[]{} Sommer-Larsen J., Gelato S. \& Vedel H., 1999, ApJ, 519, 501
\bibitem[]{} Somerville R. S. \& Primack J. R., 1999, MNRAS, 310, 1087
\bibitem[]{} Springel V., 1999, MNRAS, {312}. 859
\bibitem[]{} Steinmetz M. \& M\"uller E., 1995, MNRAS, 276, 549
\bibitem[]{} van der Kruit P.C., 1987, A\&A, 173, 59
\bibitem[]{} Vedel H., Hellsten U. \& Sommer-Larsen J., 1994, MNRAS, 271, 743  
\bibitem[]{} Warren M. S., Quinn P. J., Salmon J. K. \& Zurek W. H., 1992, ApJ, 399, 405
\bibitem[]{} Weil M. L., Eke V. R. \& Efstathiou G., 1998, MNRAS, 300, 773 
\bibitem[]{} Xu K., 1998, Gas-Kinetic Schemes for Unsteady Compressible Flow Simulations, VKI report 1998-03 von Karmann Institute Lecture Series
\bibitem[]{} Xu K. \& Prendergast K. H., 1994, J. Comput. Phys., {114}, 9
%\bibitem[]{} Xu K., Martinelli L.,  Jameson A., 1995, J. Comput. Phys., {120}, 48
%\bibitem[]{} Xu K., Kim L., Martinelli L., Jameson A., 1996, J. Comput. Phys., {7}, 213
\bibitem[]{} Yoshii Y. \& Sommer-Larsen J.,  1989, MNRAS, {236}, 779

\end{thebibliography}
\end{document}